\documentclass[nologo,11pt,a4paper,hidelinks]{ETHpaper}

\title{Social nucleation: Group formation as a phase transition}
\author{Frank Schweitzer\footnote{Corresponding author: \texttt{fschweitzer@ethz.ch}}, Georges Andres}
\authoralternative{Frank Schweitzer, Georges Andres}
\address{Chair of Systems Design, ETH Zurich, Weinbergstrasse 58, 8092 Zurich, Switzerland}
\reference{(accepted for publication)}

\usepackage{graphicx, amsmath, amssymb}
\usepackage[numbers,sort&compress]{natbib}
\usepackage{enumitem}
\usepackage{xcolor}
\setlist{nolistsep}
\usepackage{makecell}
\usepackage{booktabs}

\newcommand{\mean}[1]{\left\langle #1 \right\rangle}
\newcommand{\abs}[1]{\left| #1 \right|}
\renewcommand{\epsilon}{\varepsilon}

\begin{document}
\maketitle

\begin{abstract} The spontaneous formation and subsequent growth, dissolution, merger and competition of social groups bears similarities to physical phase transitions in metastable finite systems.
  We examine three different scenarios, percolation, spinodal decomposition and nucleation, to describe the formation of social groups of varying size and density. 
  In our agent-based model, we use a feedback between the opinions of agents and their ability to establish links.
  Groups can restrict further link formation, but agents can also leave if costs exceed the group benefits.
  We identify the critical parameters for costs/benefits and social influence to obtain either one large group or the stable coexistence of several groups with different opinions.
  Analytic investigations allow to derive different critical densities 
  that control the formation and coexistence of groups.
Our novel approach sheds new light on the early stage of network growth and the emergence of large connected components.

\end{abstract}

\section{Introduction}
\label{sec:introduction}

Models of social systems frequently utilize the network approach where
nodes represent individuals, or agents in general, while links represent social interactions between individuals.
This approach is widely used not only in social physics \citep{weidlich-book-00,castellano2009,Schweitzer2018,BATTISTON20201}, but also in the social sciences which have established their own tradition to study social networks already back in the 1940s \citep{scott1988social,prell2012social}.
It advocates a \emph{structural} perspective where nodes can be characterized based on the topological properties of the network.
The different centrality measures to quantify the importance of nodes are good examples to demonstrate the success of this approach \citep{friedkin1991theoretical,everett2005extending,bonacich}.
At the same time, they also illustrate the limitations of the structural perspective.
The temporal sequence of interactions is neglected \citep{Scholtes2016,falzon2018embedding}, despite the fact that the bursty nature of the social dynamics is important \citep{Karsai-Bursty}.
Further, the internal dynamics of nodes play no role \citep{macy2002factors,Schweitzer2020}.
If agents represent social individuals, internal degrees of freedom cannot be ignored \citep{bonabeau2002agent,M_s_2013,macy2002factors}.
Individuals make decisions, for example to whom to establish a relation or when to leave a group, they consider costs and benefits before joining, e.g., an online social network.
Hence, links between individuals are not primarily established by chance, but by choice.
The \emph{Why} matters as much as the \emph{How}.

These issues become of importance when the \emph{formation} of networks shall be explained.
Established model classes from the social sciences, for instance exponential random graph models (ERGM) \citep{lusher2013exponential,cranmer2011inferential}, aim at including social mechanisms such as homophily, reciprocity  or triadic closure \citep{currarini2016simple,gorski2020homophily,block2019forms,Schweitzer2019} to explain the formation of social links.
The problems of these models to cope with repeated interactions, the temporal order of interactions, or simply with a large number of agents, cannot be discussed here \citep{leifeld2018temporal,krivitsky2011adjusting}.

But models proposed in the context of statistical physics or random graph theory do not fare better.
They mostly feature even simpler mechanisms of random link formation,
like the random graph models from the 1950s \citep{erdos_1960,gilbert_1959} or the small world network model \citep{watts1998collective}. 
The probability of link formation can be also biased to take into account, e.g. the preferential attachment to nodes with a high degree or a high (static) ``fitness'' \citep{golosovsky2018mechanisms,musmeci2013bootstrapping}, or to nodes with similar degree (assortativity) \citep{noldus2015assortativity,catanzaro2004social}.
Such models have the problem, in addition to the \emph{ad hoc} motivation of the attachment rules, that they necessarily lead to a largely \emph{connected} network characterized by a certain degree distribution.
That is understandable because their focus is on the \emph{result}
rather than the \emph{process} of network formation.

Our aim is to change this focus toward the early stages of network growth.
To adequately capture this dynamics, we develop a new perspective that combines processes of social group formation with physical models of phase transitions.
Our approach considers that large connected social systems can emerge from  different mechanisms \citep{Fronczak2007,rivera2010dynamics,PODOBNIK2019613,Capocci2006,arenas2008motif,Singh2016}.
Online social networks are the most studied ones \citep{musial2013creation,bielenberg2012growth,mislove2008growth,garcia2013b,Capocci2006}.
They grow when new users join and link to established users .
The costs involved are low, therefore we 
observe a high entry rate, and users create many links.
This often results in a large connected component and a core-periphery structure of the social network \citep{BATTISTON20201,Garc_a_Mu_iz_2006,Gamble_2015}.
Disconnected components also exist, but they are comparably small and only contain a negligible fraction of users.

In the offline social world it is more costly to establish links.
It is already costly to find the right partners, relations need to be maintained with more effort and will be discarded if not beneficial \citep{Schweitzer2019c,Collins2018}.
Hence, instead of one large connected network, we find a number of different groups coexisting \citep{Onnela2011,Brodka2013}.
Individuals are densely connected within, but not necessarily across groups \citep{granovetter_1973,BATTISTON20201,bianconi2014triadic}.
Individuals can leave a group to join another one, groups can also merge if they find that they have enough in common.
Further, groups can polarize \citep{macy2003polarization,flache2017,schweighofer_2020_2,schweighofer_2020} because they represent different cultures or opinions, they can compete for members that have to be convinced to join \citep{Boyd2009,M_s_2013,groeber_2009_2}.

In order to model these processes of group formation from a unifying perspective, we need to consider a number of social ``ingredients'', which are explained in more detail in Section~\ref{sec:basics}.
Agents in our model have to overcome an entry barrier if they want to form a new group or join an existing one.
They should consider costs and benefits of belonging to a group \citep{Wilkins2010,Schweitzer2019c}. 
Groups can influence their members and can build up a group identity. 
Based on this, groups should have the ability to restrict the admission of agents.
In addition to the formation of one large group, we should also allow for the stable coexistence of different groups, even for competition between groups.

To achieve such a unifying perspective, we build on one central feedback.
In our model, agents are characterized by an internal continuous variable, generally speaking an ``opinion'', which can change over time.
Their opinion determines the possibility to establish links to other agents, this way forming a group.
The group, on the other hand, influences the ability of its members to create further links.
Hence, there is a feedback between \emph{opinion dynamics} and \emph{group formation}.
With additional assumptions about socially motivated costs and benefits and about the boundary conditions for network formation we are able to develop a large variety of group structures as demonstrated in Section~\ref{sec:clust-form-netw}. 

The physical models of phase transitions come into play when we try to distinguish \emph{three different scenarios} of group formation and their ability to establish one large group.
For this classification, we develop analogies between group formation 
of individuals and phase transitions in metastable and unstable thermodynamic systems.
The spontaneous formation of a new phase under supercritical conditions is known from spinodal decomposition \citep{cahn1965phase,langer1971theory} and from  percolation in porous media \citep{Stauffer_1979}.
Nucleation processes, on the other hand, first lead to a larger number of clusters of subcritical size and only a few may spontaneously grow, to form the new thermodynamic phase \citep{abraham1974homogeneous, Book-Ulbricht1988,oxtoby1992homogeneous, schmelzer2005nucleation}.  

From these three scenarios, so far only percolation has been discussed in the context of network formation \citep{Shao2009, Allard2018,Eguiluz2003,solomon-et-00, Callaway2000, schweitzer_2021}.
When gradually adding links and nodes to a network, 
percolation describes in the emergence of a giant connected component, which resembles a second-order phase transition.
Recently, models for explosive percolation have been proposed that allow for new universality classes in the characterization of such phase transitions.
These models introduce new mechanisms, such as the product rule \citep{achlioptas_2009}, to influence the type of phase transitions.
But nucleation or spinodal decomposition as established mechanisms of thermodynamic phase transitions have not been considered to describe the formation of networks.

We will close this gap in our paper, which is organized as follows:
In Section~\ref{sec:basics}, we recap some basics of thermodynamic phase transitions, to provide the concepts later used in the paper.
The main part of this section is devoted to introduce, and to formalize, the social components of our model.
We will also demonstrate how the restrictions for agents to form links impact the percolation threshold as one measure of a phase transition.

In Section~\ref{sec:clust-form-netw}, we introduce a stochastic approach to group formation by motivating transition rates for different processes of growth and dissolution of groups.
We then present a large number of agent-based computer simulations to illustrate the three different scenarios.
Eventually, we provide a systematic study of the parameter space, to distinguish the three scenarios.

Section~\ref{sec:analyt-invest} presents analytical investigations of our model and corresponding computer simulations, to fully understand the growth, dissolution, competition and coexistence of groups.
By analyzing two limit cases, incremental growth and densification, we are able to derive formal expressions for critical densities that capture the essential differences in the dynamics of groups.

In Section~\ref{sec:discussion-1}, we spend effort to assemble the various analytical and simulation results into a comprehensive and coherent view of group formation as a phase transition.
We then link the discussion back to our starting point, by examining the relevance of our results for modeling social systems.
Eventually, we provide insights of how our modeling approach shall be applied and extended toward multidimensional opinion dynamics and multi-layer networks.
It manifests that our modeling approach sheds new lights on social processes of group formation, and has the potential to open new routes 
to study the dynamics of social networks.

\section{From physical to social models}
\label{sec:basics}

\subsection{Kinetics of phase transitions}
\label{sec:mech-phase-trans}

\paragraph{Control parameter. \ }

As a reference point for our discussions, we shortly summarize kinetic aspects of phase transitions in thermodynamic systems \citep{Book-Ulbricht1988,oxtoby1992homogeneous,schmelzer2005nucleation}.
Consider a finite system in a gaseous state with the boundary conditions $N,V,T$ = const.
$N$ is the number of molecules in a vapor, for example, $V$ is the system volume and $T$ is the temperature.
The corresponding thermodynamic potential is the free energy $F(N,V,T)$.
For fixed $N$, $V$ the temperature determines the stability of the gaseous system.
Specifically, we can define a density $\rho=N/V$ and an equilibrium density $\rho^{\mathrm{eq}}(T)$.
Then the control parameter $y=\rho/\rho^{\mathrm{eq}}(T)$, known as \emph{supersaturation}, describes whether the system is in an equilibrium, $y=1$ or an unstable state, i.e. $y\gg 1$.
In the  supersaturated case, we can observe a phase transition, for instance the formation of small water droplets in a water vapor.
Whether a macroscopic liquid phase is formed, crucially depends on the size of the system and the value of the supersaturation \citep{09-fs-lsg-87}.
System size matters because of the conservation of molecules.
If droplets are formed, we have a thermodynamic system with two phases, the gaseous phase $(\beta)$ and the liquid phase $(\alpha)$.
Hence
\begin{align}
  \label{eq:30}
  N=&N_{0}(t)+\sum_{k=1}^{K}n_{k}(t) N_{k}(t)
\end{align}

where $N_{0}$ is the number of molecules in the gaseous phase, while the sum contains the number of molecules in the liquid phase. 
$n_{k}$ is the size of a droplet, given by the number of molecules, and $N_{k}$ is the number of droplets with size $n_{k}$.
Thus, once droplets are formed, the initial supersaturation $y$ decreases, and we have instead the \emph{actual} supersaturation $y_{0}(t)=N_{0}(t)/[V\rho^{\mathrm{eq}}(T)]$.

Additionally, there are energetic considerations.
Assume spherical droplets with a radius $r_{k}=[3n_{k}/(4\pi\rho_{\alpha})]^{1/3}$, where $\rho_{\alpha}$ is the density of the liquid phase.
These droplets are characterized by a surface tension $\sigma$.
The formation of this surface requires the Gibbs surface energy $W=(4\pi\sigma/3)r_{k}^{2}$, while the formation of the bulk phase releases energy \citep{schmelzer2006classical}.
This results in a critical droplet radius
\begin{align}
  \label{eq:31}
  r^{\mathrm{cr}}=\frac{2\sigma}{C}\frac{1}{\ln y}
\end{align}
where $C$ is a dimensionality constant that contains the density $\rho_{\alpha}$ and the temperature. 
Droplets that have reached the critical size will grow further until the free energy has reached its minimum. 
But to obtain the critical size requires to overcome the energy barrier characterized by a maximum of the free energy and needs to consider fluctuations \citep{abraham1974homogeneous,17-fs-88}.
\paragraph{Three scenarios of phase transitions. \ }

Eqn.~\eqref{eq:31} already points to the conditions under which phase transitions occur.
We can use these conditions to distinguish between three different scenarios. 
Firstly, we can have a very large initial supersaturation $y$, which characterizes an unstable system.
Then the energy barrier is negligible and the macroscopic liquid phase forms immediately, surrounded by the saturated gaseous phase. 
This is known as \emph{spinodal decomposition} \citep{cahn1965phase,langer1971theory}.

Secondly, we can have a medium initial supersaturation, but a negligible
value of the surface tension $\sigma$.
Then, again, the energy barrier is negligible, and a macroscopic liquid phase forms until the system saturates.
This dynamics has analogies to \emph{percolation}, where links are formed between occupied lattice sites \citep{Stauffer_1979}.
It is known that for a 2d regular lattice of size $N$ the critical density is $\rho^{\mathrm{cr}}=n_{c}/N=0.593$, i.e. if $n_{c}$ lattice sites are occupied, we can expect to find a percolating cluster in the limit of large $N$, which is the equivalent of a macroscopic phase. 
There is no surface tension involved in the formation of clusters.

The third scenario, \emph{nucleation}, is the most interesting one.
It is characterized by a medium initial supersaturation and a non-negligible surface tension.
This implies a rather large energy barrier and thus a large critical radius.
Hence, the system is initially in a \emph{metastable} state.
This results in the formation of a larger number of small  droplets of subcritical size.
Dependent on the supersaturation, a fraction of these droplets can reach a supercritical size and grow further.
This reduces the actual supersaturation drastically and no new droplets can form.
But the supercritical droplets still have to form a macroscopic phase.
This dynamic process is known as Ostwald ripening \citep{voorhees1985theory,marqusee1983kinetics,12-schmelzer-fs-87-b,Book-Ulbricht1988}.
Established droplets can grow further only if:
\begin{align}
  \label{eq:32}
  \frac{dr_{k}(t)}{dt}& =\frac{2\sigma}{C} \left[\frac{1}{r_{0}^{\mathrm{cr}}(t)}-\frac{1}{r_{k}(t)}\right]
\end{align}
Here, $r_{0}^{\mathrm{cr}}(t)={2\sigma}/{[C\ln y_{0}(t)]}$ is the actual critical radius that depends on the actual supersaturation $y_{0}(t)$.
Droplets with a subcritical radius shrink via re-evaporation, which allows droplets with a supercritical radius to further grow. 
Additionally, processes of \emph{coagulation} can be considered which  happen if droplets of different size collide.
Also processes of \emph{fragmentation} can occur, but for droplets they are negligible.

\subsection{Group formation}
\label{sec:mech-group-form}

How can we link the three different scenarios for phase transitions summarized above to the formation of social groups?
Just renaming droplets as groups will not lead to any new insights,
but would be also wrong.
We shortly discuss the main differences in the following.

\paragraph{Groups with varying density. \ }

Firstly, social groups are not spherical clusters with radius $r_{k}$, equal bulk density $\rho_{\alpha}$ and equal surface tension $\sigma$.
They are rather like small \emph{networks} (see Figure~\ref{fig:symbolic representation}), characterized by the number of group members  which we call \emph{agents} in the following, and the number of links that connect them.
Connected agents form a \emph{group} $g_{k}[n_{k},m_{k}]$, where $k$ is the group \emph{index}, $n_{k}$ is the number of agents and $m_{k}$ is the number of links in group $k$.
$K$ then denotes the \emph{total number} of groups, which can change over time. 

We further define a group density $\rho_{k}=2m_{k}/n_{k}$.
It  is not a constant either, but can change by agents joining or leaving or by adding or removing social links.
The factor of 2 reflects that each link connects two agents, so we normalize the number of links to $n_{k}/2$. 
Note that $\rho_{k}$ is normalized to the \emph{size} of the group rather than to the number of possible links, $n_{k}(n_{k}-1)$,  as used in network science. 
Indeed, our definition is akin to the average degree of an agent in the group.

The minimum number of links in a group is $m_{k}^{\mathrm{min}}=(n_{k}-1)\simeq n_{k}$, i.e. each agent is connected via only one link, for example in a star-like topology.
Hence $\rho_{k}=2n_{k}/(n_{k}-1)\simeq 2$  if only \textit{incremental growth} is considered, i.e. groups grow by incorporating one new agent at a time through the formation of a single link.
Note that $\rho_{k}=2$ is the limiting density for large groups in the case of incremental growth.
Small groups always start from two agents with one link, i.e. $\rho_{k}=1$.

On the other hand, the maximum number of (undirected) links in a group is $m_{k}^{\mathrm{max}}=n_{k}(n_{k}-1)/2\simeq n_{k}^{2}/2$, i.e. agents form a fully connected network.
Then $\rho_{k}^{\mathrm{max}}=(n_{k}-1)\simeq n_{k}$ is the maximum density
if groups are allowed to densify, i.e. agents create links within a group. 
With these considerations it holds that  $\rho_{k}\in [1,\{2, n_{k}\}]$.

\paragraph{Group utility. \ }

To further distinguish social groups from droplets, we characterize groups of different $n_{k}$, $m_{k}$ by means of a \emph{utility}:
\begin{align}
  \label{eq:88}
  u_{k}[n_{k},m_{k}] = b m_{k} - c n_{k}
\end{align}
It reflects the social insight that being part of a group has benefits and costs.
The benefits $bm_{k}$ arise from the existence of links within a group.
More links allow more interactions, a better exchange of information, etc., as argued in management science \citep{Uzzi-network-effect}.
This differs from definitions of utilities in, e.g., economics that assign maintenance \emph{costs} to links \citep{Battiston2012}.
On the other hand, maintaining a group is costly, therefore 
$cn_{k}$ denotes a cost proportional to the number of agents that need to be integrated into the group.
The linear form for the utility function is chosen because of its minimal assumptions.
We have no evidence for suitable nonlinear forms.
Further, we do not make the utility dependent on local substructures of the group that differ across agents, but on the group as a whole. 

Once groups are established, agents will consider costs and benefits from being part of a group and will leave if not satisfied, as explained later in Sect.~\ref{sec:transition-rates}.

The \emph{maximum} utility is $u_{k}^{\mathrm{max}}=(b/2)\, n_{k}^{2} - c n_{k}$.
The \emph{minimum} utility, on the other hand, is $u_{k}^{\mathrm{min}}=(b-c)n_{k}$.
Thus, if $(b-c)<0$, $u(n_{k},m_{k})=0$ defines the minimum size of a group to have a positive utility, namely $\hat{n}_{k}=(b/c)m_{k}$.
Using the group density $\rho_{k}=2m_{k}/n_{k}$, 
the respective density is then: $\hat{\rho}=2c/b$.
This allows us to rewrite the utility as
\begin{align}
  \label{eq:24}
  u_{k}(n_{k},\rho_{k}) = c\, n_{k}\left[\frac{\rho_{k}}{\hat{\rho}}-1 \right]
\end{align}
As we will show below, 
groups with $\rho_{k}> \hat{\rho}$ likely remain or even grow, while groups with $\rho_{k}< \hat{\rho}$ likely dissolve.

\paragraph{Homophily. \ }

In addition to the varying densities $\rho_{k}$ and utilities $u_{k}$, there are dynamic peculiarities that distinguish social groups from droplets.
The social principle of \emph{homophily} \citep{mcpherson2001birds,mcmillan2022worth,gorski2020homophily,currarini2016simple} states that agents tend to interact more often with those others that are similar in some respect.
To formalize this, we characterize agents by a \emph{scalar variable} $x_{i}$, drawn from a uniform distribution $U(0,1)$.
$x_{i}(t)$ could represent, in a very general sense, the \emph{opinion} of agent $i$ with respect to some issue, which can change over time as will be discussed below.
The uniform distribution of initial opinions is chosen because it requires the least constraints and because it  allows to consider more extreme opinions, which are likely to be underrepresented by a normal distribution.

Two randomly chosen agents $i$ and $j$ will successfully interact only if the difference in their $x$ values is below a certain \emph{threshold} $\epsilon$, which is also interpreted as a \emph{tolerance} for deviating opinions.
To formalize this, we introduce a variable:
\begin{align}
  \label{eq:19}
  z_{ij}(t)= \epsilon - \abs{x_{j}(t)-x_{i}(t)}
\end{align}
Agents successfully interact only if $z\geq 0$.
Their interaction triggers another feedback, \emph{convergence}: Because of \emph{social influence} \citep{mavrodiev2013quantifying,flache2017,groeber_2014,flache2011local} agents' opinions tend to become more similar if they interact more \citep{Helbing2011crowd, mavrodiev_2021}.
To include a dynamics for $x_{i}(t)$, we adopt for simplicity the bounded confidence model \citep{deffuant_2000,hegselmann_2002,lorenz_2007,groeber_2009_2}.
A more complex opinion dynamics for multidimensional opinions can be considered as well \citep{schweighofer_2020,schweighofer_2020_2}.
Two randomly chosen agents $i$ and $j$ update their opinions $x_{i}$, $x_{i}$ \emph{if} they interact as follows:
\begin{equation}
  \label{eq:mean}
  \frac{\Delta x_{i}(t)}{\Delta t}= \gamma \left[x_{j}(t) - x_{i}(t)\right] \Theta[z_{ij}(t)]
\end{equation}
Here $\mathrm{\Theta}[z]$ is the Heaviside function, which returns 0 if $z<0$ and 1 if $z\geq 0$. 
If we choose the maximum value $\gamma=0.5$, there is convergence to the mean $(x_{i}+x_{j})/2$ in one time step.
Smaller values of $\gamma$ require more interactions to obtain more similar opinions. 
We note that the dynamics of Eqn.~\eqref{eq:mean} together with a random sequential update of agents' opinions is \emph{path dependent}, i.e. the sequence of interactions matters for determining the final opinion of an agent. 

\paragraph{Group formation and group influence. \ }

Our main assumption is that group formation builds on successful interactions.
That means, if  two randomly chosen agents $i$ and $j$ are able to interact because their opinions are similar enough, $z_{ij}>0$, they will also form a link, $a_{ij}=1$: 
\begin{align}
  \label{eq:19a}
a_{ij}(t)=\mathrm{\Theta}[z_{ij}]
\end{align}
This is the central mechanism for the formation of groups:
while interactions can happen randomly, 
the formation of links depends on similarity, i.e. the partner is not randomly chosen.
This considers that  social interactions are costly.
We emphasize that the existence of a link indicates a \emph{special relation} between agents, which is to be distinguished from a mere interaction. 
Therefore, once agents formed a group with other agents of similar opinions, they tend to keep these relations.   
This implies that groups continue to have an influence on the opinion of their members \citep{holyst2001social}, which is captured  in an \emph{effective opinion} \citep{groeber_2009_2}:
\begin{align}
x_i^{\textrm{eff}}(t) =& \left[1-\alpha\right]x_i(t) + \alpha \mean{x_i^k(t)}
                   \label{eq:13}
\end{align}
$ \mean{x_i^k(t)}=\sum_{j\in \abs{n_{k}}} x_{j}(t)/n_{k}$
is the average opinion of group $k$, and $\alpha$ weights its influence on agent $i$.
Considering the group influence, we can modify $z_{ij}$ in Eqn.~\eqref{eq:19} as follows: 
\begin{align}
  \label{eq:33}
  z_{ij}(t)= \epsilon - \abs{x^{\mathrm{eff}}_{j}(t)-x^{\mathrm{eff}}_{i}(t)}
\end{align}
That means, instead of the individual opinion of an agent, its weighted opinion determines whether a link is formed or deleted.
In the limit $\alpha=1$ the effective opinion for everyone in the group is the same.

We emphasize that group influence reflects another social mechanism, \emph{triadic closure} \citep{klimek2013triadic,bianconi2014triadic,Schweitzer2019}.
Open triads refer to three agents $i$, $j$ and $k$ which are connected by only two links $(i,j)$ and $(j,k)$.
Then the probability that either $i$ or $k$ form the third link $(i,k)$  to close the triad is much larger than the probability that either $i$ or $k$ form a link to an agent $r$ outside the group.
Without group influence, these probabilities would only depend on $x_{i}$, $x_{k}$ and $x_{r}$.
With group influence, $x_{i}^{\mathrm{eff}}$ and $x_{k}^{\mathrm{eff}}$ become more similar, so the likelihood that the link is formed inside the group increases.
This process will lead to the \emph{densification} of the group to improve triadic closure.

\subsection{Network formation}
\label{sec:basic-mechanisms}

\paragraph{Percolation threshold. \ }

We now change the perspective from the group level to the system level, which can be described as a large social network.
The $a_{ij}$ defined in Eqn.~\eqref{eq:19} are the entries of a symmetric adjacency matrix $\mathcal{A}$ of size $N\times N$, which captures the topology of the network.

A group is a \emph{disconnected component} of this network.
Initially, social groups may form spontaneously in a random  manner.
Therefore in an early stage of the network evolution we may expect a larger number of disconnected components. 
The number of these disconnected components, their sizes and their  densities can grow and shrink over time. 
A phase transition is characterized by the emergence of a \emph{giant connected component}, i.e. one group of macroscopic size that includes most of the agents.

To better understand the conditions under which we can expect the formation of a giant connected component in the network, let us turn to one of the simplest network models, known as the $G(N,M)$ model.
It requires to fix the number of agents, $N$, and the total number of links, $M$.
Initially there is no network.
It is formed successively by randomly choosing, out of $N$ agents, two agents $i$ and $j$ and connect them by a link. 
Each established link diminishes the number of available links by one.
The process is repeated until all $M$ links are spent.
The result is a \emph{random network} with characteristic properties that can be described analytically \citep{Bollob_s_2001}.

The $G(N,p)$ model looses the assumption of a fixed $M$ \citep{gilbert_1959}.
Instead, two randomly chosen agents are connected with a certain probability $p$, which is determined such that it matches the \emph{expected degree} $\mean{d}$, i.e. the expected number of links.
If $N^{2}/2$ is the maximum number of links between $N$ agents and $M$ the total number of links, we then find for $p$:
\begin{align}
  \label{eq:18}
  p=\frac{2\mean{M}}{N^{2}}\;; \quad \mean{d}=\frac{2\mean{M}}{N}=pN
\end{align}

The question whether we will observe a giant connected component from this simple procedure depends on the total number of links, $M$, which has to have a critical value, known as the percolation threshold.
Because we do not assume any underlying geometry, such as lattices, the percolation threshold results from the condition $Np=1$ \citep{molloy_1995,schweitzer_2021}.
Using the relations from Eqn.~\eqref{eq:18}, it follows directly that a percolating group can be expected if $M^{\mathrm{cr}}=N/2$.

This allows us to define an initial supersaturation for \emph{links} $\mu=M/M^{\mathrm{cr}}$.
To expect a phase transition via percolation, we have to assume that
 $\mu=\mu^{\mathrm{cr}}=1$.
This result holds for infinite systems.
For finite systems, this value has to be larger as we will show in Sect.~\ref{sec:percolation}.
Nevertheless, $\mu^{\mathrm{cr}}=1$ is a good reference value to distinguish whether a system has the ability of a phase transition.
We could compare it to the supersaturation introduced in Sect.~\ref{sec:mech-phase-trans}, $y=\rho/\rho^{\mathrm{eq}}(T)$, which has the same interpretation, just in a thermodynamic context.

\paragraph{Restricted link formation. \ }

The above discussion holds as long as agents are not restricted in their link formation.
In our model, however, we have introduced a threshold $\epsilon$, Eqn.~\eqref{eq:19}, which determines whether two agents $i$ and $j$ can interact, i.e. can potentially form a link. 
So far $\epsilon$ was large enough to allow link formation between \emph{all} agents. 
Now we consider that not all agents can create links to all other agents, i.e. $\epsilon\ll 1$.

To still ensure a phase transition, we have to calculate the critical  value of $\epsilon$ needed to allow for a sufficient number of links.
For this we follow an earlier approach \citep{schweitzer_2021} that links percolation in random networks with so-called \emph{threshold networks}, i.e. networks where the link formation is restricted by the tolerance threshold $\epsilon$.

We remind that the $x_{i}$ are uniformly distributed in the interval $[0,1]$ and that for the formation of links $\Delta x_{ij}=\abs{x_{j}-x_{i}}$ matters.
The distribution function $P(\Delta {x})$ of the absolute difference between two uniform variables is given by the \emph{triangular distribution}.
The cumulative probability $F(\Delta {x} \leq \epsilon)$ to find a value $\Delta {x} \leq \epsilon$ follows likewise:
\begin{align}
  \label{eq:5}
  P(\Delta {x}) =&\ 2 -2 \left(\Delta {x} \right) \;; \quad 0\leq \Delta {x} < 1 \nonumber \\
F(\Delta {x} \leq \epsilon) = &\ 2 \epsilon - \epsilon^{2}
\end{align}
$F(\epsilon)$ is the probability that for a randomly chosen pair of agents a link is \emph{possible}.
This has to be multiplied by the independent probability that this pair of agents is also chosen to form a link, which is given by $p$.
Hence, the condition for percolation $Np=1$ has to be corrected by the factor $F(\epsilon)$ that a link formation could not be possible if opinions deviate.
With this consideration and Eqn.~\eqref{eq:18} we find:
\begin{align}
  \label{eq:34}
Np\,F(\epsilon)=\frac{2\mean{M}}{N}\left[2\epsilon -\epsilon^{2}\right]  
\end{align}
From this quadratic equation for $\epsilon$ 
only the lower value makes sense in our model because  $\epsilon$ has to have values between 0 and 1.
With $\mu=2M/N$, it solves into:
\begin{equation}
  \label{eq:3}
  \epsilon^{\mathrm{cr}} = 1 - \sqrt{1 - \frac{1}{\mu}}
\end{equation}
$\epsilon^{\mathrm{cr}}$ allows us to choose the tolerance for deviating opinions such that still a giant connected component can emerge, dependent on the link supersaturation $\mu$.  
If $\mu=\mu^{\mathrm{cr}}=1$, i.e. we just have the conditions for the onset of percolation in an infinite system, $\epsilon=1$, i.e. we cannot allow for restrictions in the interactions.
The higher the link supersaturation, the lower $\epsilon$ can be to still observe a giant connected component.

For $\epsilon< \epsilon^{\mathrm{cr}}$, we can only expect the formation of disconnected components.
This is similar to the bounded confidence model, where  the value of $\epsilon$ determines the final number of groups that can each reach consensus.
For this number, a rough estimate $1/(2\epsilon)$ was found. 
That means one group, i.e. a giant connected component, can be only obtained for $\epsilon\geq 0.5$.
$\epsilon\leq 0.25$ would already result in two groups, i.e. two disconnected components \citep{deffuant_2000,lorenz_2007}.

In our case, for $\mu=3$, for instance, we find a giant connected component for $\epsilon\geq 0.2$.
This comparably low value becomes clear, because the giant connected component is a ``spanning cluster'', i.e. it can be a very sparse network, as long as it spans the whole system.
In Sect.~\ref{sec:clust-form-netw}, 
we will illustrate how such giant connected components look like.

We note that Eqn.~\eqref{eq:3} is derived under the assumption of uniformly distributed $x_{i}$.
It does not consider a change of this distribution at the same time scale as the formation of the giant connected component.
Hence, in order to apply Eqn.~\eqref{eq:3} we have to assume that $\gamma$ in Eqn.~\eqref{eq:mean} is considerably small.
In fact, in Figure~\ref{fig:perc} we later use $\gamma$=0.1, and we further show in Figure~\ref{fig:time} that in the percolation scenario the formation of the largest group occurs very fast.

\section{Simulating group formation in networks}
\label{sec:clust-form-netw}

\subsection{Transition rates}
\label{sec:transition-rates}

In the following we lay out a stochastic dynamics for the formation of groups.
As the discussion in Sect.~\ref{sec:mech-phase-trans} made clear, we need to consider fluctuations in order to describe the spontaneous formation of groups and their growth up to a critical size.
The dynamics of groups with a supercritical size, on the other hand, can be described by a deterministic dynamics, which will be developed in Section~\ref{sec:analyt-invest}.

The fundamental dynamics of stochastic processes is described by a master equation.
This can be defined at different levels.
The most explicit perspective starts from the adjacency matrix that contains all information about links between agents.
Our focus is \emph{not} on the description of the full network, but on the dynamics of individual groups.
This leaves us still with the choice between the group level or the full group distribution, $\pmb{N}=[N_{1,0},N_{2,1},N_{3,2},N_{3,3},...,N_{N,M}]$.
The elements $N_{n,m}$ count the number of groups of size $n$ with $m$ links.
This approach is summarized in appendix B.
In the following, we focus on the group level because we do not depend on the full description of the dynamics of the state space.
In fact, all our equations can be derived from the simpler perspective.
Hence, 
in our case we start from $p[g_{k}(n_{k},m_{k},t)]$, the probability to find group $k$ with $n_{k}$ agents and $m_{k}$ links at time $t$.

The specification of the dynamics requires to define the respective \emph{transition rates} for the change of these probabilities.
This can be done in different ways.
Let us first discuss the perspective of \emph{agents}.
To determine the formation of a link $(i,j)$ between two randomly chosen agents $i$ and $j$, we have to consider the probabilities (i) that agents $i$ and $j$ are chosen, which is $1/N$ for each in a random sampling, and (ii) that they are able to establish  a link which depends on the value of $z_{ij}$, Eqn.~\eqref{eq:33}, and (iii) that links are still available. 

$M$ is the initial number of links which is limited because we distinguish between interactions and links.
The latter are only formed under certain conditions which are socially motivated by the opinion dynamics already discussed.
Social networks are sparse, therefore, $M$ has to be smaller to avoid fully connected networks.
Similar to the $G(N,M)$ model, the fixed $M$ also determines the probability of link formation.

If links were unrestricted, the system could evolve to maximal density always.
In real social systems, interactions can be abundant, but social relations, i.e. links, are usually limited, e.g. by mental capacities.  We remind on the Dunbar number.
The restriction is further given by the limited size of the groups.
We are considering finite systems with $N$ agents, hence the maximum number of relations would be $M=N(N-1)/2$.
If we would go with this number, the model would end up with fully connected networks.

But at time $t$ some groups have already been formed.
Therefore we have
\begin{align}
  \label{eq:1}
  M=m_{0}(t) +\sum_{k=1}^{K}m_{k}(t)\;;\quad
   N=n_{0}(t) +\sum_{k=1}^{K}n_{k}(t)
\end{align}

$m_{0}(t)$ denotes the number of free links and $n_{0}(t)$ the number of free agents that do not belong to any group at time $t$.
With these assumptions the transition rate for link formation reads:
\begin{align}
  \label{eq:20}
  w\left[(i,j)|i,j\right]\propto \frac{2 m_{0}}{N^{2}}\Theta[z_{ij}(t)] 
\end{align}
In the limit of $\epsilon\to 1$, $\Theta[z]=1$.
We note that this rate holds no matter whether agents are already part of a group or not. 

We can describe the same process from the \emph{group} perspective.
Each group is characterized by two values $n_{k}$, $m_{k}$.
When we randomly choose two agents, the probability that one of the agents is in group $k$ and the other one in group $l$ is proportional to $n_{k}$ and $n_{l}$.
With these consideration  the transition rate for a growing group reads in general:

\begin{align}
  \label{eq:21}
  w\left[(n_{k}+n_{l},m_{k}+m_{l}+1)|(n_{k},m_{k}), (n_{l},m_{l})\right] = w^{+}\left[g_{k},g_{l}\right] \propto \frac{2 m_{0}}{N^{2}}
  \sum_{i\in g_{k}}\sum_{j\in g_{l}}\Theta[z_{ij}(t)] 
\end{align}

\begin{figure}[htbp] \includegraphics[width=0.99\linewidth]{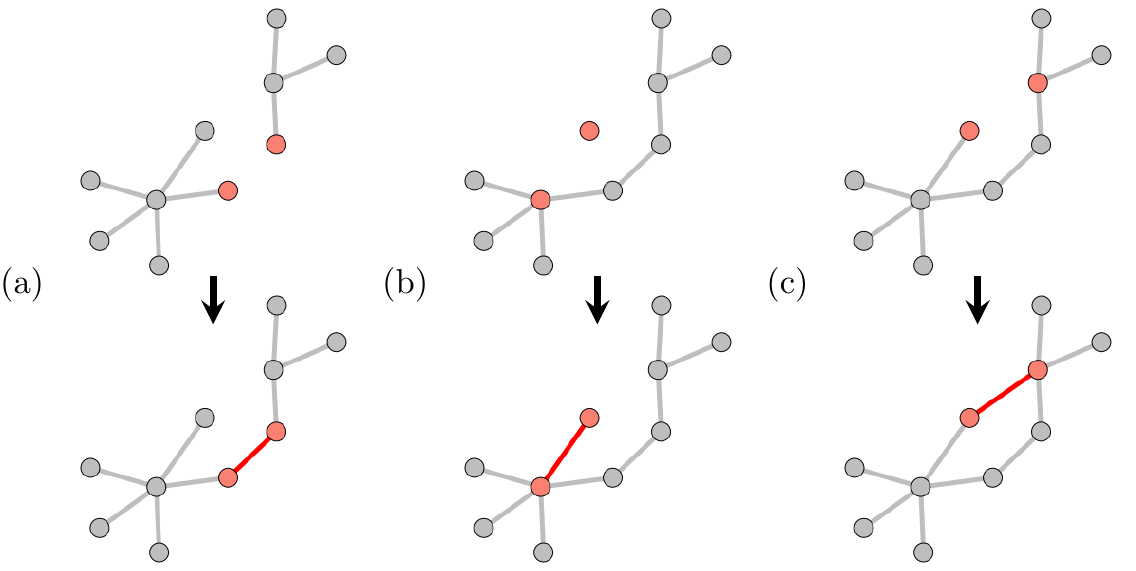}
 \caption{Illustration of the three growth mechanisms for groups: (a) coagulation, Eqn.~\eqref{eq:21}, (b) incremental growth, Eqn.~\eqref{eq:21-inc}, (c) densification, Eqn.~\eqref{eq:14}.}
  \label{fig:symbolic representation}
      \end{figure}

In the limit of $\epsilon\to 1$, $\Theta[z]=1$ the summation term becomes $n_{k}n_{l}$.
This rate describes the process of \emph{coagulation}.
Adding one link between two groups $g_{k}$, $g_{l}$ leads them to merge, i.e. to form one connected group of size $(n_{k}+n_{l},m_{k}+m_{l}+1)$.
This can be seen as a jump in the $(n,m)$ space, as illustrated in Figure~\ref{fig:symbolic representation}(a).

If an agent $i$ has not joined a group yet, it defines its own group, i.e. then $n_{i}$ = 1 and $m_{i}$ = 0 and $g_{i}(n_{i},m_{i})=g_{1}(1,0)$.
In this case, Eqn.~\eqref{eq:21} captures the   process of  \emph{incremental growth} and reads specifically:

\begin{align}
  \label{eq:21-inc}
  w\left[(n_{k}+1,m_{k}+1)|(n_{k},m_{k}), (1,0)\right] = w^{+}\left[g_{k},g_{1}\right] \propto \frac{2 m_{0}}{N^{2}}
  \sum_{i\in g_{k}}\sum_{j\in g_{1}}\Theta[z_{ij}(t)] 
\end{align}

Adding one link between a group $g_{k}(n_{k},m_{k})$ and an isolated agent, $g_{1}(1,0)$ grows both $n_{k}$ and $m_{k}$ by 1, incrementally, as shown in Figure~\ref{fig:symbolic representation}(b).

Additionally, there can be also a  \emph{densification} of the group, illustrated in Figure~\ref{fig:symbolic representation}(c). 
If the two agents already belong to the same group, adding one link does not change the number of agents, but only the number of links and this way the density.
The new group is then given by $g_{k}(n_{k},m_{k}+1)$.
In general, the transition rate for densification reads:

\begin{equation}
  \label{eq:100}
  w\left[(n_{k},m_{k}+1)|(n_{k},m_{k})\right]
  \propto  \frac{2 m_{0}}{N^{2}}\sum_{\substack{i,j \in n_{k} \\ \{i,j\}\not\in m_{k}}}\delta_{ij} \Theta[z_{ij}]
\end{equation}

The sum over the Kronecker $\delta_{ij}$ counts only 
agents $i$ and $j$ that both belong to the group of $n$ agents, but do not have a common link  yet.
If we assume $\epsilon=1$, i.e. $\Theta[z_{ij}]=1$, 
the probability $p_{ij}$ that such a link already exists can be calculated from the configuration model as:

\begin{equation}
  \label{eq:11}
  p_{ij} = \frac{d_{i}d_{j}}{2{m_{k}}}
\end{equation}

$d_{i}$ and $d_{j}$ are the degrees of agents $i$ and $j$, hence the product considers all possibilities to connect them.
This has to be normalized by $m_{k}$, the existing number of links in the group $k$.

We further know that in a group of density $\rho_{k}$ each agent has on average $\mean{d}_{k}=2m_{k}/n_{k}=\rho_{k}$ links.
We use this to replace $d_{i}$ and $d_{j}$ in Eqn.~\eqref{eq:11}.
Hence, the sum in Eqn.~\eqref{eq:100} is expressed by the probability that agents $i$ and $j$ are chosen to form a link, which is proportional to $n_{k}^{2}$, multiplied by the independent  probability that they do not already have one:
\begin{equation}
  \label{eq:120}
  p_{ij} \equiv p_{k} = \frac{\rho_{k}^{2}}{2m_{k}} = \frac{2m_{k}}{n_{k}^{2}} \;;\quad
  \sum_{\substack{i,j \in n_{k} \\ \{i,j\}\not\in m_{k}}}\delta_{i,j} =
  n_{k}^{2}\left[1-\frac{2m_{k}}{n_{k}^{2}} \right]
\end{equation}
This  allows to rewrite, for $\epsilon=1$, the transition rate for \emph{densification}, Eqn.~\eqref{eq:100} as:

\begin{align}
  \label{eq:14}
  w\left[(n_{k},m_{k}+1)|(n_{k},m_{k})\right] =
  w^{+}\left[ g_{k},m_{0}\right] &\propto  \frac{2 m_{0}}{N^{2}}n_{k}
  \left[n_{k}-\rho_{k} \right]
\end{align}

The difference $\omega_{k}=n_{k}-\rho_{k}$ tells how many links could be potentially added to group $k$.
Hence, this transition rate decreases with higher density.
Furthermore, when the group is fully connected, we have $m_{k} = {n_{k}(n_{k}-1)}/{2}$. This means that in the limit of large systems ($n_{k}(n_{k}-1)\rightarrow n_{k}^{2}$) no new link can be formed, which is reasonable.
For finite systems this holds only approximately.

With this, we have considered three different processes for network \emph{growth} via the formation of groups. 
But we also need to specify how groups can dissolve \citep{Carro2014}.
For this, we have already proposed in Sect.~\ref{sec:mech-group-form} that agents leave a group if they experience an negative utility, i.e. if costs exceed benefits.
According to Eqn.~\eqref{eq:24}, this implies that for a group $g_{k}$, it holds that $\rho_{k}<\hat{\rho}$.
We propose that agents leave a group \emph{spontaneously} at the following rate:

\begin{align}
  \label{eq:400}
w\left[(n_{k}-1,m_{k}-\delta m_{k}) | (n_{k},m_{k})\right] = w^{-}\left[g_{k},\rho_{k}\right] \propto \frac{n_{k}}{N}  \exp{\left\{\beta\frac{\hat{\rho}}{\rho_{k}}\right\}}
\end{align}

$n_{k}/N$ is again the probability that an agent from group $g_{k}$ is randomly chosen to leave.
The exponential term ${\hat{\rho}/\rho_{k}} = (c n_{k})/(b m_{k})$
can be seen as some sort of inverse fitness of a group.
If it is larger than one, which means costs are larger than benefits, the probability that the group dissolves spontaneously increase exponentially.
$\beta$ is the inverse temperature, $1/T$, and defines the level of randomness. 
Smaller $\beta$ make differences between benefits and costs more important, larger $\beta$ smooth out this influence.

If an agent leaves a group, all its links in the group are removed and become available as free links.
Thus, the group size $n_{k}$ is deminished by 1.
But how many links will be removed?
We know that in a group of density $\rho_{k}$ each agent has on average $\mean{d}_{k}=\rho_{k}$ links.
That means, the expected change is $\delta m_{k}=\rho_{k}$ in Eqn.~\eqref{eq:400}.
We note that this transition rate in principle also describes the \emph{fragmentation} of a group of size $n_{k}$ into pieces.
In this case just the arguments of $w^{-}$ have to be changed, the probability for a spontaneous leave remains the same.

We summarize, that the fastest way for a network to \emph{grow} is by coagulation. 
Once some groups have formed, they can merge into much larger groups.
This helps to speed up a possible phase transition.
If coagulation is excluded, i.e. $n_{l}\equiv 1$, the phase transition would occur much slower because incremental growth is a rather slow process.
So we could see it as the  \emph{worst case} scenario.
Densification, on the other hand, does not lead to growth,
but \emph{stabilizes} existing groups because it increases the group utility.
But even if the density is increased, as long as $\rho_{k}<\hat{\rho}$, staying in such a group would not be beneficial.
Hence, groups with a smaller density are likely to shrink and to disappear, no matter what their size $n_{k}$ is.

\subsection{Results of agent-based computer simulations}
\label{sec:results-agent-based}

Using the above defined transition rates, we can write down $k$ master equations \citep{Roepke_2013} for the growth and dissolution of groups.
For the probability $p[g_{k}(n_{k},m_{k},t)]$ the stochastic dynamics reads  in general:

\begin{align}
  \label{eq:8}
  \frac{\Delta p[g_{k}(n_{k},m_{k},t)]}{\Delta t}=& \sum\nolimits_{g_{l}} w^{+}\left[g_{k},g_{l}\right]\;  p[g_{k}(n_{k}-n_{l},m_{k}-m_{l},t) \cap g_{l}(n_{l},m_{l},t)]  \nonumber \\
   & - w^{+}\left[g_{k},g_{l}\right]\;  p[g_{k}(n_{k},m_{k},t) \cap g_{l}(n_{l},m_{l},t)]  \nonumber \\
                                        &  + w^{+}\left[g_{k},m_{0}\right] \; p[g_{k}(n_{k},m_{k}-1,t)]  - w^{+}\left[g_{k},m_{0}\right]\;  p[g_{k}(n_{k},m_{k},t)] \nonumber \\
                                                  &    + w^{-}\left[ g_{k},\rho_{k}\right] \; p[g_{k}(n_{k} + n_{l},m_{k} + m_{l},t) \cap g_{l}(n_{l},m_{l}, t+\Delta t)] \nonumber \\
                                                 &   -  w^{-}\left[ g_{k},\rho_{k}\right] \; p[g_{k}(n_{k},m_{k},t)]
\end{align}

Note that the master equations for the $k$ different groups are coupled via the transition rates that have to satisfy the boundary condition, Eqn.~\eqref{eq:1}.
To simulate the dynamics we use the Gillespie algorithm.
That means, for each time step we calculate all possible transition rates.
Their relative weight is used to randomly select one of the possible transitions.
In the following we present simulation results for the three different scenarios for phase transitions described in Sect.~\ref{sec:mech-phase-trans}.
We decided to keep these physical labels for the scenarios even for the dynamics of social groups, simply to allow for comparison. 

\paragraph{Percolation. \ }
\label{sec:percolation}

\begin{figure}[htbp] \includegraphics[width=0.28\textwidth]{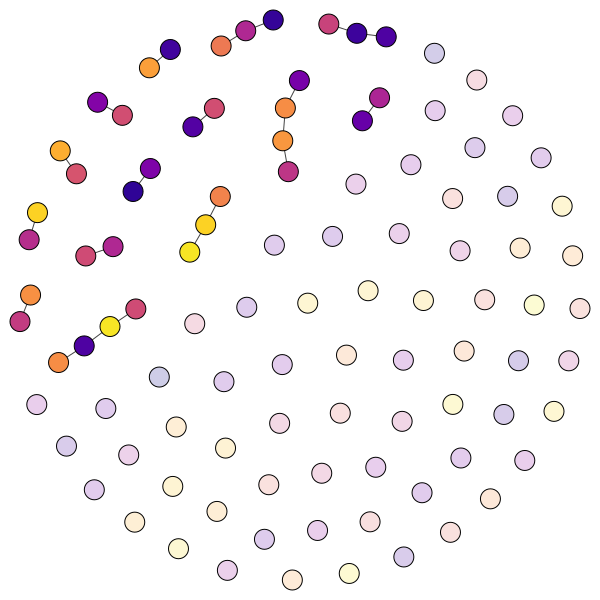}
  (a) \hfill
   \includegraphics[width=0.28\textwidth]{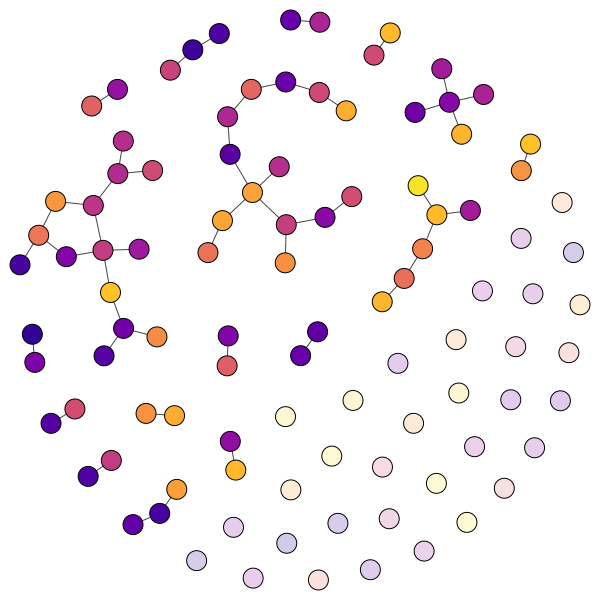}
  (b) \hfill
   \includegraphics[width=0.28\textwidth]{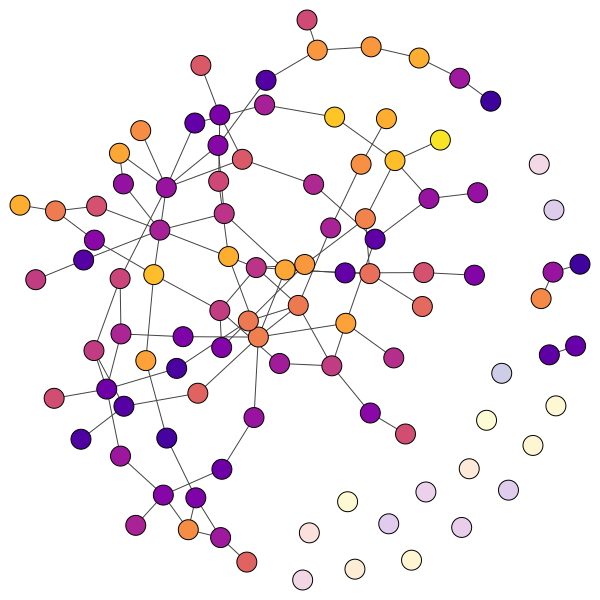}
  (c)
  \caption{Phase transition via percolation. Different time steps: (a) $t$ =  0.14, (b) $t$ = 0.55, (c) $t$ = 1.42. Parameters: $\mu/2$ = 1.6, $c$ = 0, $\epsilon$ = 1, $\gamma$ = 0. The color scheme indicates the $x_{i}$ values of the agents, which do not change.}
  \label{fig:perc}
\end{figure}

The simplest scenario of group formation assumes that agents connect randomly, this way forming a \textrm{random network}.
We do \emph{not} consider any costs or benefits for groups, further \emph{no} spontaneous dissolution is allowed.
Also, we ignore for the moment any dynamics or influence of the $x_{i}$ values in the formation of groups ($\gamma$ = 0) and link formation between all agents is allowed ($\epsilon$ = 1).  
Hence, the whole dynamics is described by the transition rate for growth, $w^{+}\left[g_{k},g_{l}\right]$, Eqn.~\eqref{eq:21}. 
In an early stage we find a larger number of disconnected components, as shown in Figure~\ref{fig:perc}(a),
These disconnected components can only \emph{grow} over time as long as there are free links available. 
Eventually, they merge to form a giant connected component, shown in Figure~\ref{fig:perc}(c). 

\begin{figure}[htbp]  \centering
  \includegraphics[width=0.45\textwidth, trim={0 0 0 2.5cm},clip]{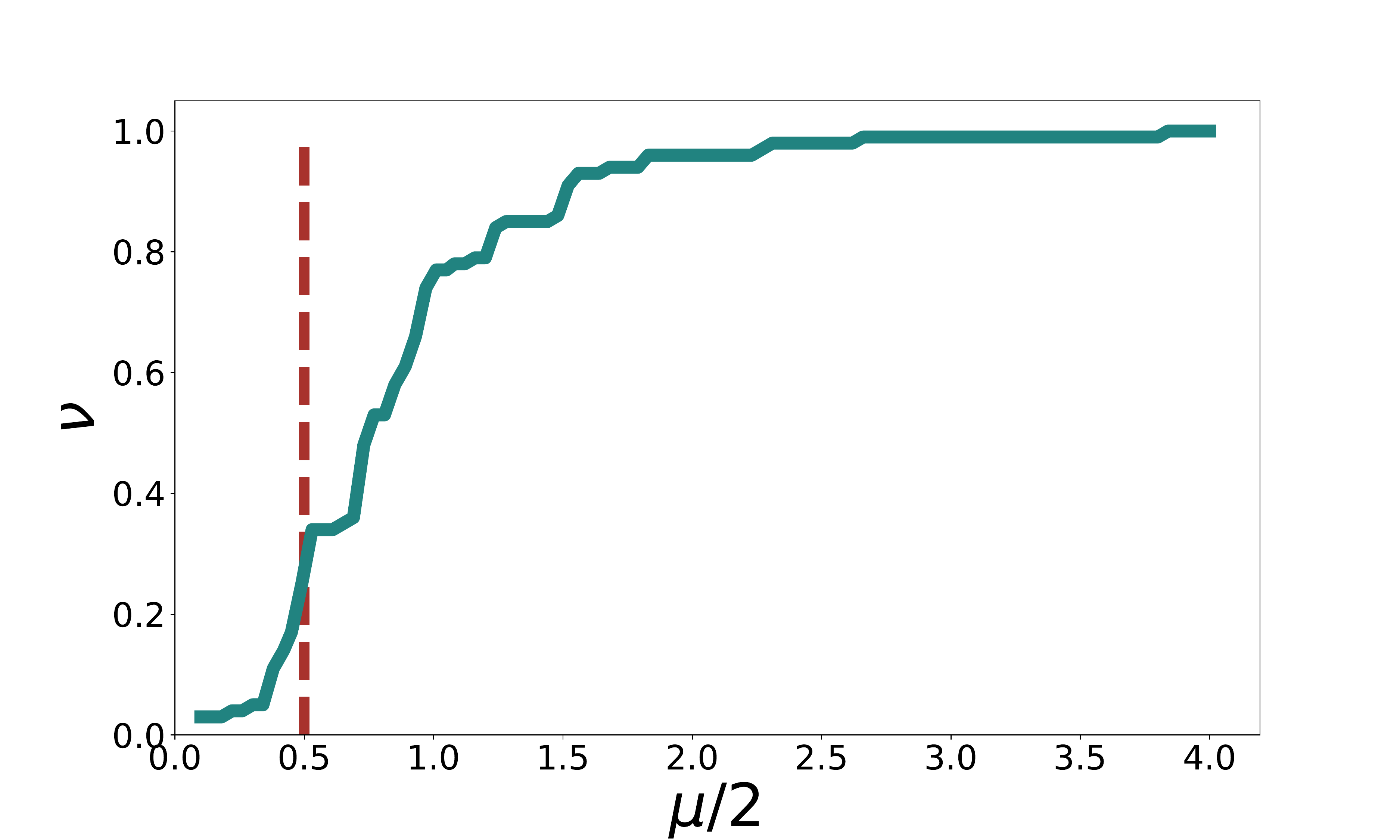}(a)
    \includegraphics[width=0.45\textwidth, trim={0 0 0 2.5cm},clip]{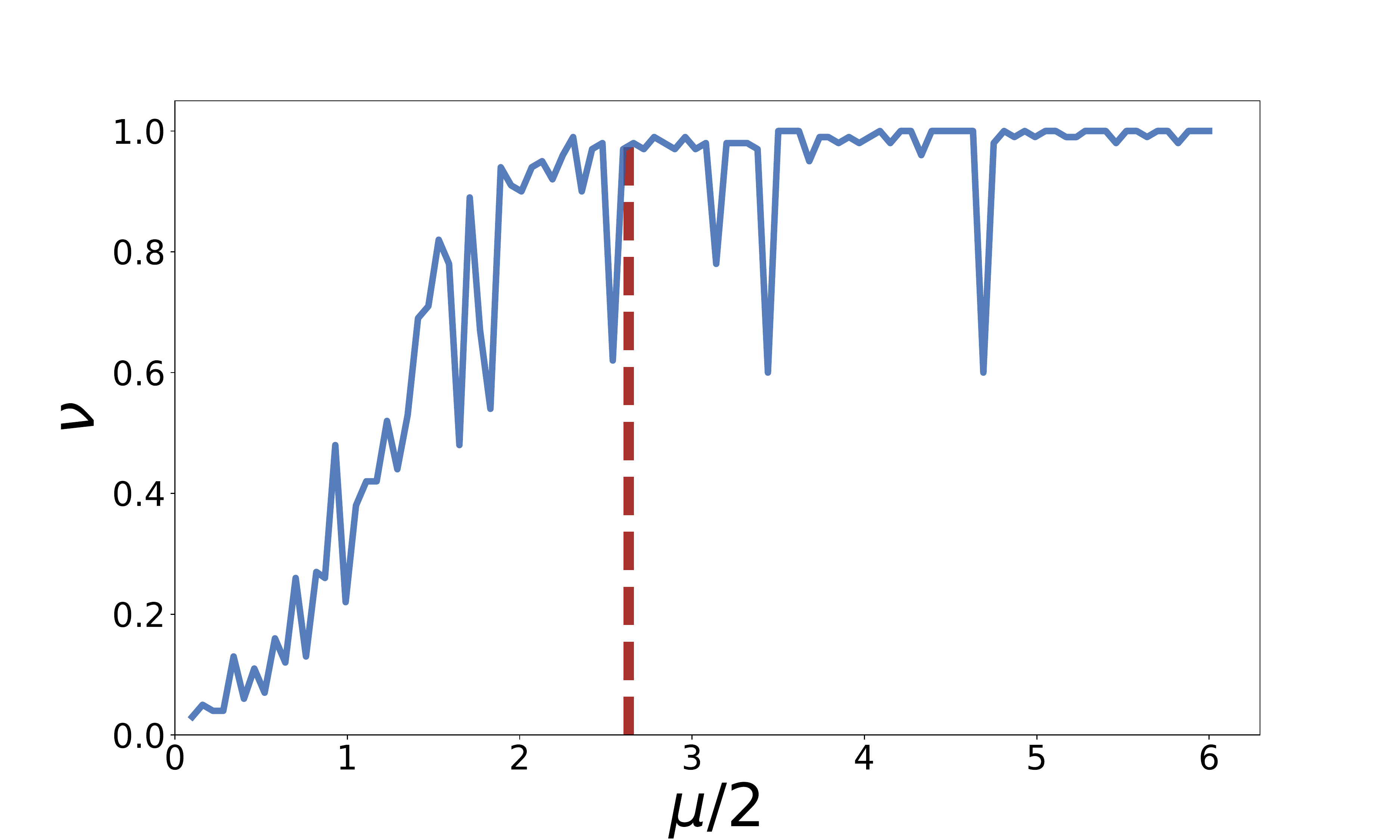}(b)
    \caption{Fraction of the largest connected group $\nu=n^{\mathrm{LCC}}/N$ for different $\mu=M/M^{\mathrm{cr}}$.
      (a)       Unconstrained interactions ($\epsilon$ = 1), (b) 
      constrained interactions $\epsilon$ = 0.1.
      The dashed lines show the theoretical predictions for the
      percolation threshold: (a) $\mu/2$ = 0.5, (b) $\mu/2$ = 2.63, Eqn.~\eqref{eq:3}.
    }
  \label{fig:thresh-pc}
\end{figure}

Figure~\ref{fig:thresh-pc} shows the fraction of the largest connected component, $\nu=n^{\mathrm{LCC}}/N$, in the final network dependent on the initial condition $\mu=M/M^{\mathrm{cr}}$, i.e. the total number of available links relative to the critical number of links that are needed for percolation.
According to the discussion in Sect.~\ref{sec:basic-mechanisms}, we expect the percolation threshold to be $\mu^{\mathrm{cr}}$ = 1, which only holds for very large systems and without any restrictions on the link formation, i.e. $\epsilon\to 1$. 
As Figure~\ref{fig:thresh-pc}(a) shows, for finite systems, $N=100$, $\mu^{\mathrm{cr}}$ only indicates the \emph{onset} of forming a percolating group.
To observe percolation with a fraction of the LCC close to one, we   need a five times larger value, as Figure~\ref{fig:thresh-pc}(a) shows. 

When restricted interactions between agents are considered, i.e. $\epsilon<1$, Eqn.~\eqref{eq:3} defines the conditions for observing a giant connected component.
For $\epsilon$ = 0.1, we find $\mu$ = 2.63.
The computer simulations shown in Figure~\ref{fig:thresh-pc}(b) give a better match with the prediction, also because in Eqn.~\eqref{eq:3} the finite number of links is explicitly considered.

We conclude that the percolation scenario is observed for  $\mu \simeq 5$, $c=0$.
Still, because the link supersaturation $\mu$ is not large, we will only obtain a \emph{a sparse percolating group}.
The initial density of free links is not large enough for a compact phase, as we show below.

\begin{figure}[htbp] \includegraphics[width=0.28\textwidth]{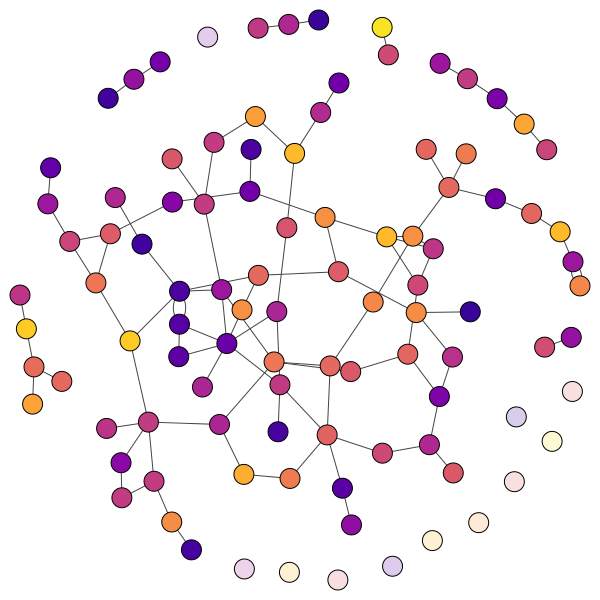}
  (a) \hfill
   \includegraphics[width=0.28\textwidth]{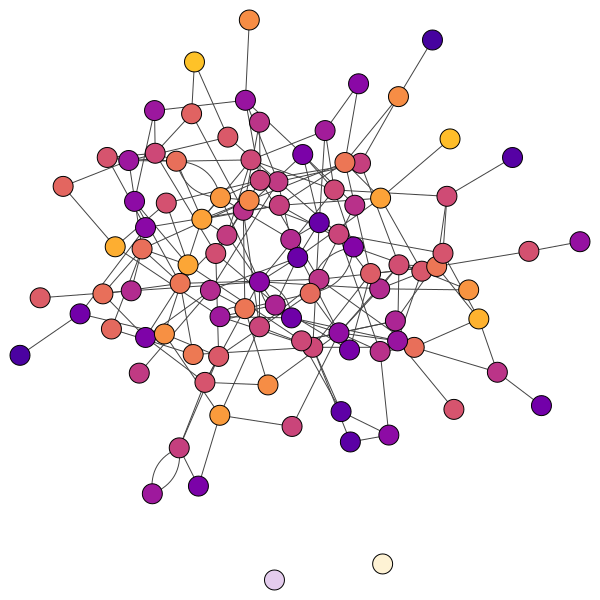}
  (b) \hfill
   \includegraphics[width=0.28\textwidth]{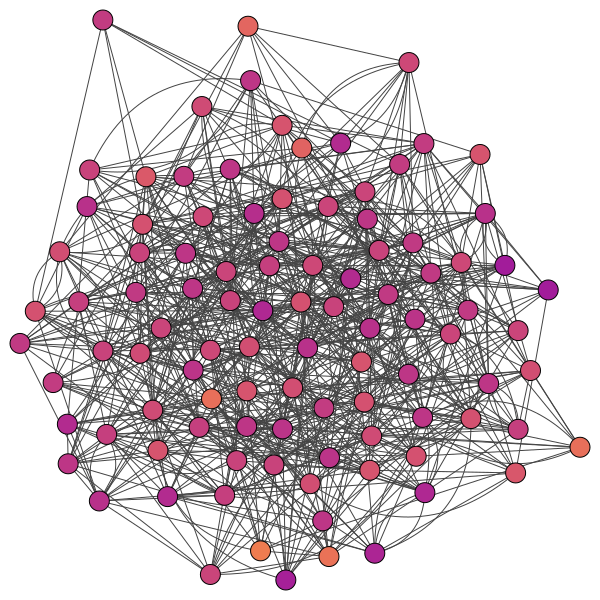}
  (c)
  \caption{Phase transition via spinodal decomposition. Different time steps: (a) $t$ = 0.08, (b) $t$ = 0.13, (c) $t$ = 0.56. Parameters: $\mu/2$ = 19, $c$ = 0.5, $N$ = 100, $\epsilon$ = 1, $\gamma$ = 0.1. The color scheme indicates the $x_{i}$ values of the agents.}
  \label{fig:spin-sim}
\end{figure}

\paragraph{Spinodal decomposition. \ }
\label{sec:spin-decomp}
As explained in Sect.~\ref{sec:mech-phase-trans} this scenario is observed if the initial system is already in an unstable state.
Further, the surface tension does not give an important effect.
In our model of group formation, this is realized by  a very large link supersaturation $\mu\gg 1$ and by a negligible cost $c\ll 1$.
Then the initial  instability is sufficient for the spontaneous emergence of a new phase, which is rather compact. 
This is illustrated in Figure~\ref{fig:spin-sim}. 

\paragraph{Nucleation. \ }
\label{sec:nucleation}

\begin{figure}[htbp] \includegraphics[width=0.28\textwidth]{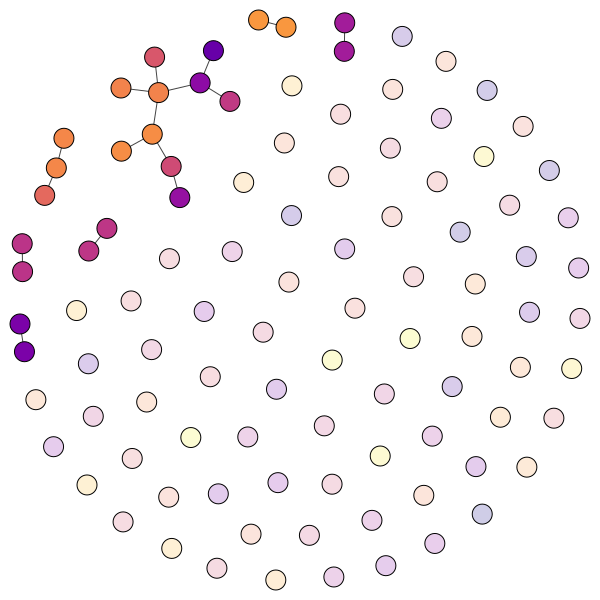}
  (a) \hfill
   \includegraphics[width=0.28\textwidth]{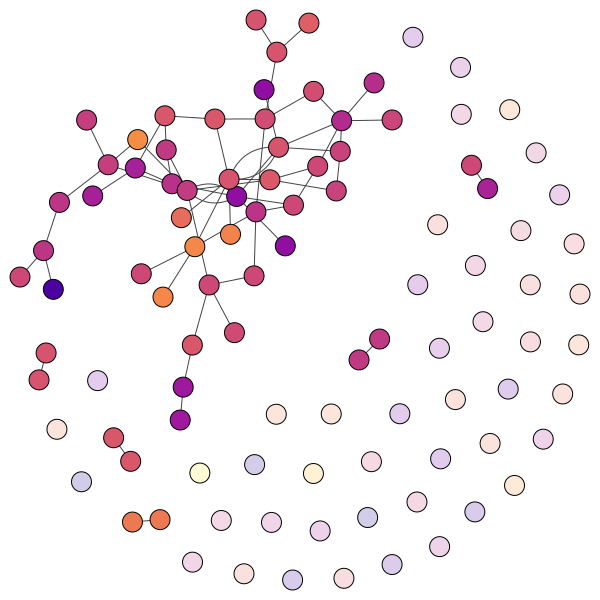}
  (b) \hfill
   \includegraphics[width=0.28\textwidth]{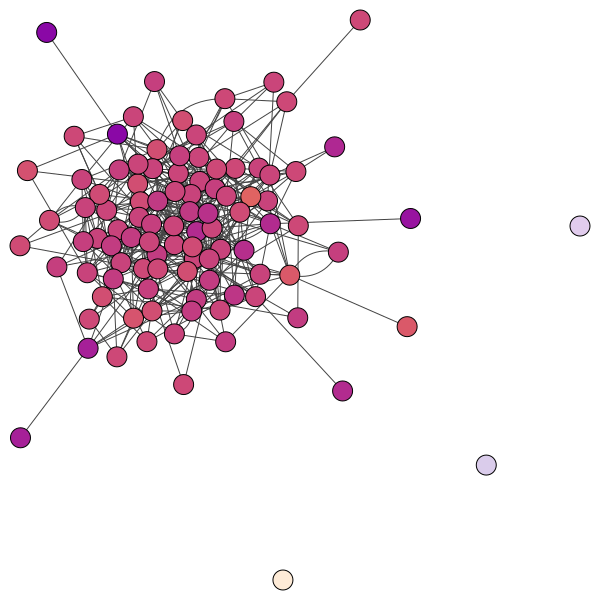}
  (c)
 \caption{Phase transition via nucleation. Different time steps: (a) $t$ = 0.7, (b) $t$ = 2.07, (c) $t$ = 2.67. Parameters: $\mu/2$ = 16, $\hat \rho$ = 15.6, $N$ = 100, $\epsilon$ = 1, $\gamma$ = 0.5, $\beta$ = 0.4. The color scheme indicates the $x_{i}$ values of the agents.}
  \label{fig:nucleation}
\end{figure}

To observe \emph{nucleation}, the initial system has to be in a clear metastable state.
In our model of group formation this is realized by a  moderate link supersaturation $\mu \gg 1$ and a non negligible cost $c>0$.
Then, the new phase can only emerge if initial fluctuations generate groups larger than a critical size, discussed in detail in Sect.~\ref{sec:crit-clust-size}. 
The phase transition occurs via the formation of a number of groups  in an early stage, shown in Figure~\ref{fig:nucleation}(a).
These group later either coalesce or they form a macroscopic phase via the redistribution of links from groups with subcritical to groups with supercritical sizes.
This process is known as Ostwald ripening and  will be further discussed in Section~\ref{sec:ostwald-ripening}.

A dynamics, where initially small groups are formed and later merge or dissolve to give way for a giant connected component, is found in the evolution of collaboration networks.
In \citep{Tomasello2017}, 
groups represent R\&D alliances in one or multiple industrial sectors over 20 years, as reconstructed from the SDC Platinum database. 
A group-size distribution akin to our intermediate regime ($t$ = $1.2$) in Figure~\ref{fig:group-dist}(a) can be found in these R\&D networks \citep{tomasello_2014}.

A similar dynamics is also observed in the evolution of scientific collaboration networks reconstructed from the Social Work Research Database \citep{eckl_2018}.
From initially disconnected groups, this system generates over time one prominent component encompassing up to 20\% of the nodes in later stages.
Many other disconnected component remain in the system to coexist with the giant component.

\paragraph{Restricted interactions. \  }
\label{sec:restr-inter}
\label{sec:constant-x_i}

\begin{figure}[htbp] \includegraphics[width=0.28\textwidth]{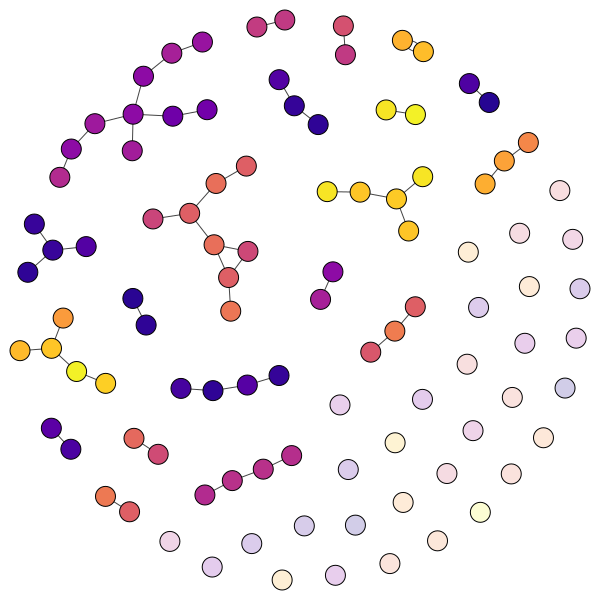}
  (a) \hfill
   \includegraphics[width=0.28\textwidth]{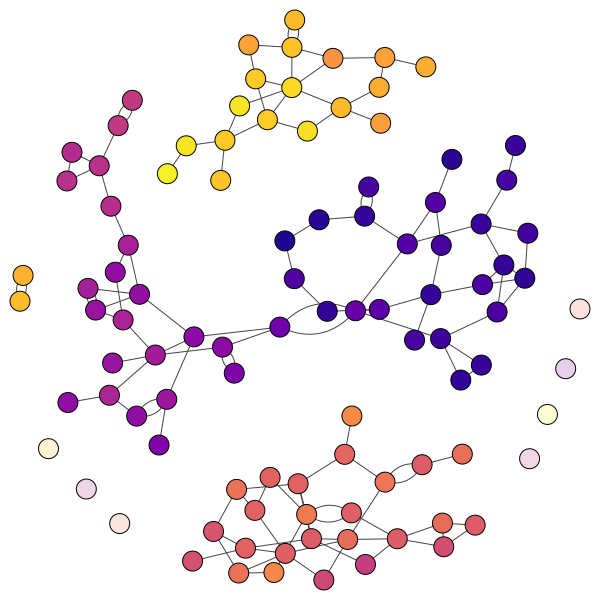}
  (b) \hfill
   \includegraphics[width=0.28\textwidth]{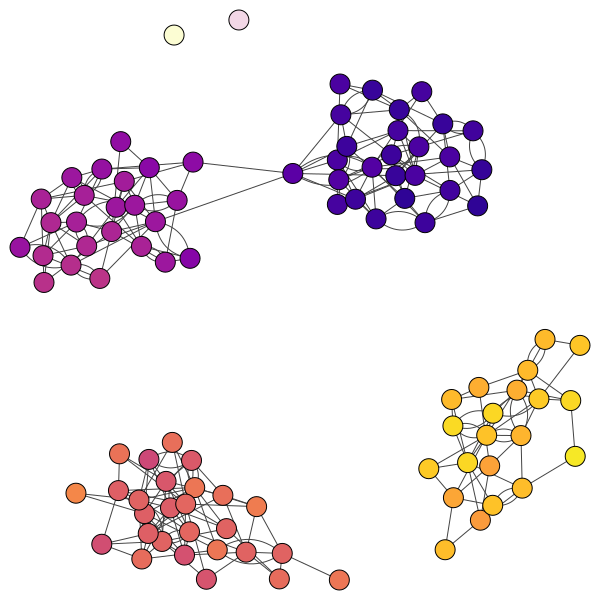}
  (c)
  \caption{Phase transition with restricted interactions. Different time steps:  (a) $t$ = 0.25, (b) $t$ = 0.68, (c) $t$ = 2.15. Parameters: $\mu/2$ = 4, $c$ = 0, $N$ = 100, $\epsilon$ = 0.125, $\gamma$ = 0.1. The color scheme indicates the $x_{i}$ values of the agents. }
  \label{fig:thres-t}
\end{figure}

As the two examples of Figures~\ref{fig:spin-sim},~\ref{fig:nucleation} show, the emerging giant connected components become rather homogeneous with respect to the opinions $x_{i}$.
This is due to the fact that all agents are allowed to form links ($\epsilon$ = 1) and the opinion dynamics, Eqn.~\eqref{eq:mean}, ensures convergence of opinions ($\gamma$ = 0.1).
If we restrict the interaction between agents by choosing $\epsilon<\epsilon^{\mathrm{cr}}$, Eqn.~\eqref{eq:3}, we observe the formation of \emph{isolated} groups instead of a giant connected component.
These groups differ considerably with respect to their average $\mean{x^{k}}$, but show rather homogeneous values of $x_{i}$ inside each group, as Figure~\ref{fig:thres-t} shows.

Again, collaboration networks show a dynamics where multiple disconnected components with non-negligible sizes survive. 
We reach sparse disconnected components  when the saturation $\mu$ is low.  
Evidence for the coexistence of sparse disconnected groups is found  for instance in the collaboration network of Italian sociologists \citep{akbaritabar_2020}.

\subsection{Parameter space} \label{sec:Phase Diagrams}

We can systematize the outcome of our agent-based computer simulations with respect to the free parameters of our group formation model.
These can be distinguished in three groups (see appendix A for table representation):
\begin{enumerate}[itemsep=0em]
\item \emph{network formation}: number of agents $N$, link density $\mu=2M/N$, level of randomness $\beta$,
\item \emph{group formation}: cost $c$, benefit $b$, or $\hat{\rho}=c/b$,
\item \emph{opinion formation}: approach rate $\gamma$, tolerance threshold $\epsilon$, group influence $\alpha$. 
\end{enumerate}
Comparing these parameters with their thermodynamic counterparts from Sect.~\ref{sec:mech-phase-trans}, we see that both $N$ and $\mu$ relate to the initial supersaturation while the cost $c$, or $\hat \rho$, relates to the surface tension, because it determines the critical group size.
The parameters of the opinion dynamics do not play a role in the thermodynamic model, because for molecules no constrains exist to form droplets.

\begin{figure}[htbp]  
  \includegraphics[width=0.99\linewidth]{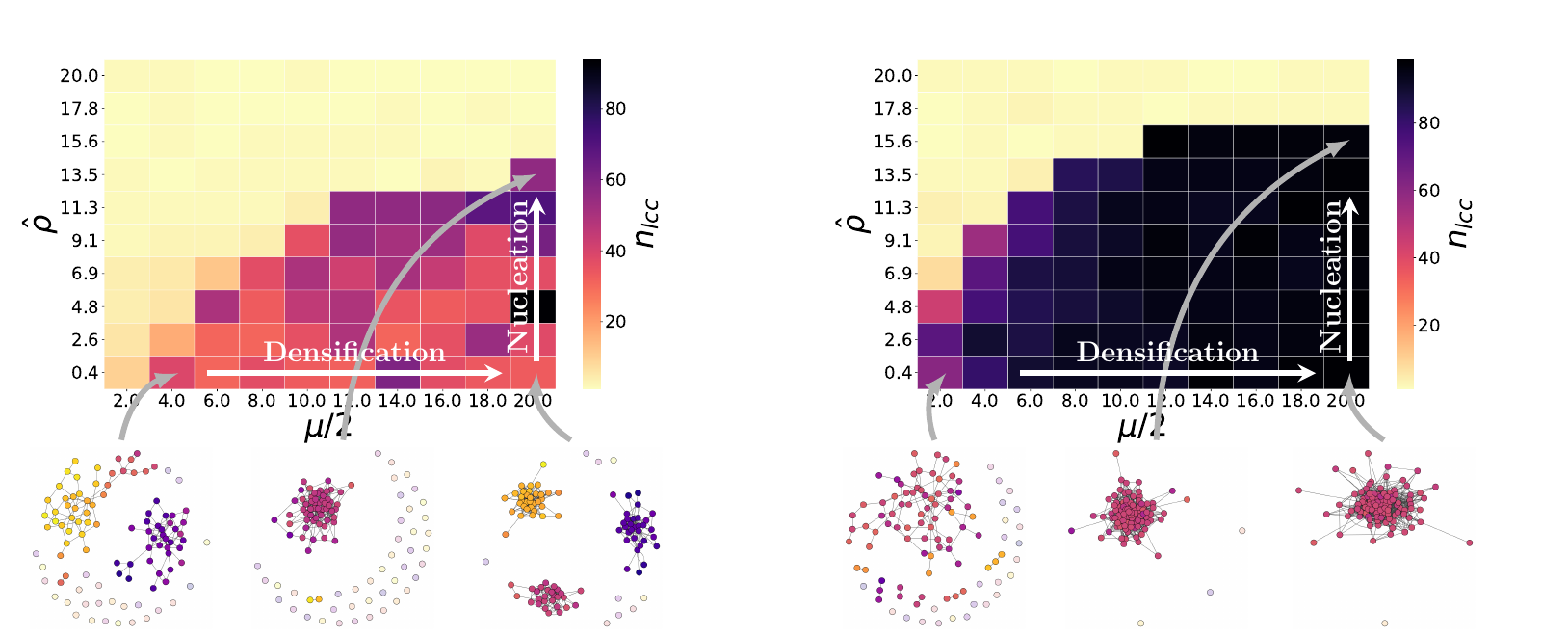}
\caption{
  $(\mu/2,\hat{\rho})$ parameter space for the size of the largest connected component $n^{\mathrm{LCC}}$. $N$ = 100, different parameters for opinion dynamics: (left) $\epsilon$ = 0.16, $\alpha$ =
0.9, $\gamma$ = 0.1, $\beta$ = 0.4, (right) $\epsilon$ = 1, $\alpha$ = 0.1, $\gamma$ = 0.5, $\beta$ = 0.4.
  }
  \label{fig: phaseDiagHighTension}
\end{figure}

In Figure~\ref{fig: phaseDiagHighTension} we present two phase diagrams of the $(\mu/2,\hat\rho)$ parameter space,  with different parameters for opinion formation (left) and (right).
These diagrams have no sharp phase boundaries, which could be only expected in the thermodynamic limit but not in finite systems.
Nevertheless, the phase diagrams succinctly illustrate the influence of the parameters on the group structure.

Our measure for the existence of a phase transition is $n^{\mathrm{LCC}}$, the size of the largest connected component (LCC), measured after 50.000 time steps, i.e. in a quasi stationary equilibrium.
With $N$ = 100, $n^{\mathrm{LCC}}$ can reach up to 100.
Comparing the left and the right diagram, we notice that an $n^{\mathrm{LCC}}$  comprising more than 90
percent of all agents is almost always obtained if the link formation between agents is not restricted by the tolerance parameter $\epsilon$ (Figure~\ref{fig: phaseDiagHighTension} right). 
Yet, if the link supersaturation $\mu$ is low, i.e. if we are close to the percolation threshold, the network of the LCC is very sparse.
It becomes more compact with increasing link saturation $\mu$.

\begin{figure}[htbp]  \centering
\includegraphics[width=0.45\textwidth]{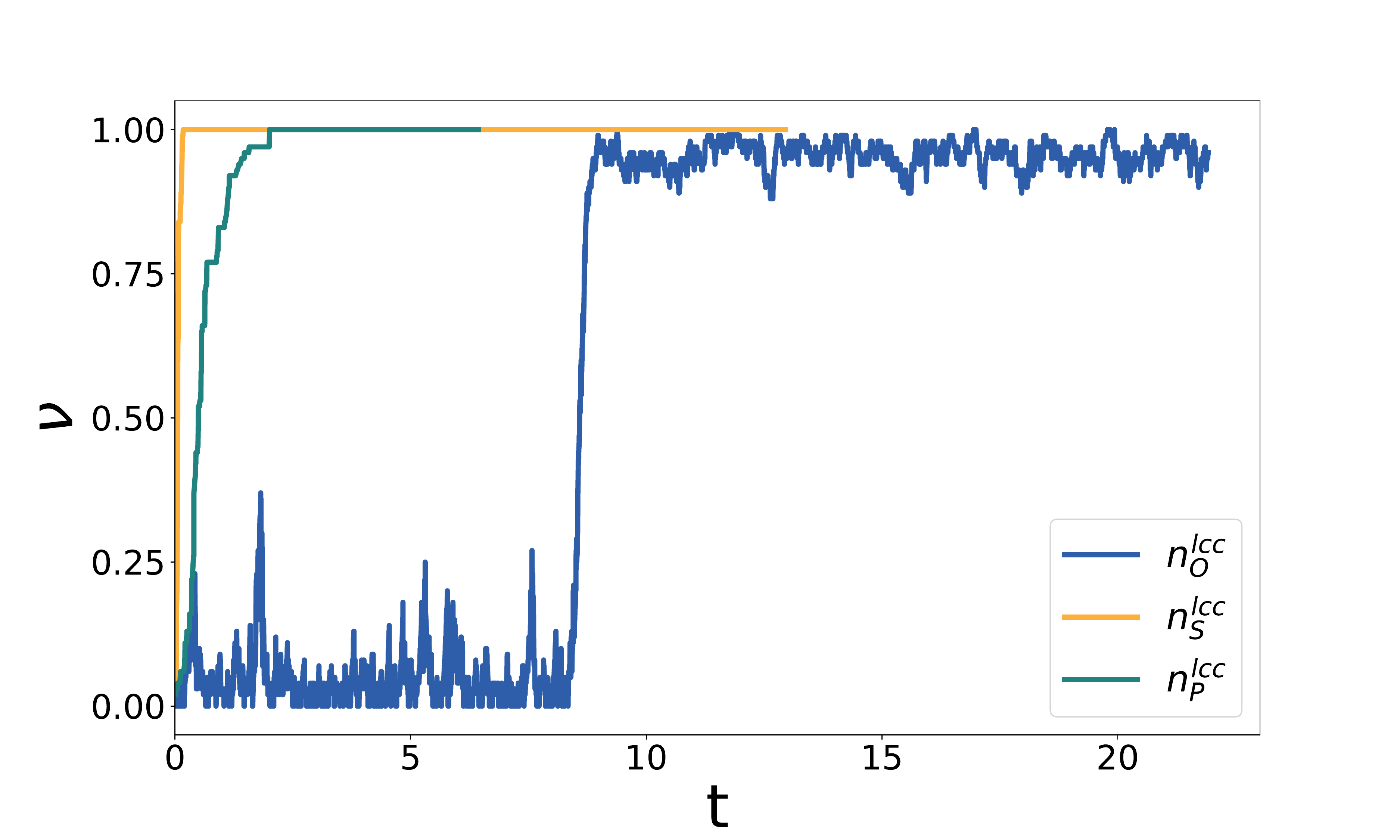}
\caption{Fraction of the largest connected group, $\nu=n^{\mathrm{LCC}}/N$, over time for the three scenarios: percolation: $n^{LCC}_{P}$ ($\mu_P/2$ = 3, $c_{P}$ = 0),  spinodal decomposition: $n^{LCC}_{S}$ ($\mu_S/2$ = 19, $c_{S}$ = 0) and nucleation: $n^{LCC}_{O}$ ($\mu_O/2$ = 19, $\hat \rho$ = 17.2). Parameters: $\epsilon$ = 1, $\gamma$ = 0.1, $\beta$ = 0.4.
      }
  \label{fig:time}
\end{figure}

In Figure~\ref{fig: phaseDiagHighTension} (right) for large $\mu$
the final state obtained for low and high $\hat \rho$ looks almost the same, however the process to reach it is very different.
For low $\hat\rho$, we are in the regime of spinodal decomposition, where one connected component is formed from the very beginning, also shown in Figure~\ref{fig:perc}.
For high $\hat \rho$ we first observe the formation of smaller groups.
The large compact group is formed by the \emph{dissolution} of the smaller ones at a much longer time scale.
This is clearly visible in Figure~\ref{fig:time}, where we plot the growth of the LCC over time for the three different scenarios percolation, spinodal decomposition and nucleation.
Both percolation and spinodal decomposition occur very fast, but for different reasons.
Percolation only considers the growth of the network, without an energy barrier or dissolution.
Spinodal decomposition, on the other hand, is driven by the very high link supersaturation, i.e. the energy barrier is negligible and dissolution does not play a role because groups formed immediately reach a supercritical size.  
Nucleation is rather slow because groups of critical size are only formed by fluctuations. 

Turning to the case of restricted link formation shown in Figure~\ref{fig: phaseDiagHighTension}(left), we observe that a giant connected component is almost never observed because agents from different groups cannot create links if the tolerance value $\epsilon$
is low, with only one exception that we explain now.

For the parameter constellation with  $(\mu/2,\hat \rho$)= (20,4.8) we observe \emph{coagulation}.
For the same $\mu$ this was not possible for smaller values of $\hat \rho$, because there large groups form quickly and the large group influence $\alpha$ makes them quite homogeneous with respect to agents' opinions.
As a consequence, these groups quickly reach average opinions too different for them to merge. 
With increasing $\hat \rho$, nucleation dominates.
More smaller groups with more different average opinions form, which increases the chance for coagulation.
Therefore, we observe the formation of a giant connected component.

For the same $\mu$, but even larger values of $\hat \rho$ the nucleation barrier rises and fewer groups are formed.
If groups establish, they reach a homogeneous opinion very fast, because of the high group influences.
This then does not allow free agents with different opinions to still join and hence the LCC becomes smaller, as we see.

Figure~\ref{fig: phaseDiagHighTension}(left) also demonstrates that the final states of  spinodal decomposition and nucleation can be very different. 
The nucleation scenario allows groups to dissolve such that agents can join larger groups.
Spinodal decomposition, on the other hand, generates only a small incentive for agents to leave groups, because of the small value for $\hat \rho$.
Therefore, the groups formed coexist in a stable state.

\section{Analytical investigations}
\label{sec:analyt-invest}

\subsection{Incremental growth}
\label{sec:crit-clust-size}

Eventually, we want to underpin our agent based simulations with some analytic results.
These can be hardly obtained if we consider the full master Eqn.~\eqref{eq:8}.
Therefore, we now discuss two limit cases, (i) incremental growth and (ii) densification.
These can be seen as the two sides of the same coin, as they both occur during a phase transition, albeit dominate at different times. 
Considering them separately allows us to better understand the driving forces of group formation and competition. 

First, we omit processes of coagulation and densification and consider only incremental growth, Eqn.~\eqref{eq:21-inc}, and dissolution, Eqn.~\eqref{eq:400}.
Let us further consider $\epsilon$ = 1, i.e. all agents are allowed to form links.
The respective growth process is illustrated in Figure~\ref{fig:symbolic representation}(b).
Precisely, a group can only grow if a new agent joins by connecting to the group with one link.
This leads to a very sparse cluster where $m_{k}=n_{k}-1$, i.e. $\rho_{k} = 2$ for large groups.
Clearly, in this limit we do not need to discuss the dynamics of $\rho_{k}$, we only focus on the size of the group, $n_{k}$. 

To find out under which conditions it is more likely to observe the growth or the dissolution of a group, we calculate the logit or \emph{odds ratio}, $G$, which is  defined as 
\begin{equation}
  \label{eq:7}
  G = \ln \frac{w^{+}\left[g_{k},g_{1}\right]}{w^{-}\left[ g_{k},\rho_{k}\right]}=
  \ln \left[\frac{2m_{0}n_{0}n_{k}}{N^{2}} \frac{N}{n_{k}\exp\{\beta \hat{\rho}/\rho_{k}\}}\right]
\end{equation}
From $G=0$ we find  the critical conditions to expect incremental group growth:
\begin{align}
  \label{eq:101}
  \rho_{n}^{\mathrm{cr}}(t)=\frac{\beta\hat{\rho}}{\ln\left[2m_{0}(t)\, n_{0}(t)/N \right]}
\end{align}
We remind that $m_{0}(t)$ gives the number of links \emph{available} at time $t$ and $n_{0}(t)$ the number of free agents, as defined in Eqn.~\eqref{eq:1}. 
If we use $\hat{\rho}=2c/b$ and define a supersaturation
\begin{align}
  y_{0}(t)=\frac{2m_{0}(t)n_{0}(t)}{N}=n_{0}(t)\mu_{0}(t) \quad \mathrm{with} \quad
  \mu_{0}(t)=\frac{2m_{0}(t)}{M^{\mathrm{cr}}}\;;\quad
    y_{0}(0)=N\frac{M}{M^{\mathrm{cr}}}\equiv y
  \label{eq:35}
\end{align}
we can rewrite the expression for the critical density very similar to the critical radius for nucleation, Eqn.~\eqref{eq:31}:
\begin{align}
  \label{eq:10}
  \rho_{n}^{\mathrm{cr}}(t)
=\frac{\beta\hat{\rho}}{\ln \mu_{0}(t) + \ln n_{0}(t) } =
  \frac{2c}{b/\beta} \frac{1}{\ln y_{0}(t)}
\end{align}
where the cost $c$ takes the place of the surface tension. 
We emphasize that $\rho_{n}^{\mathrm{cr}}$ does not make any statement about the \emph{size} of a group.
In fact it characterizes the system because it depends on the supersaturation $y_{0}(t)$ which is a system variable.

\begin{figure}[htbp] \centering
\includegraphics[width=0.45\textwidth]{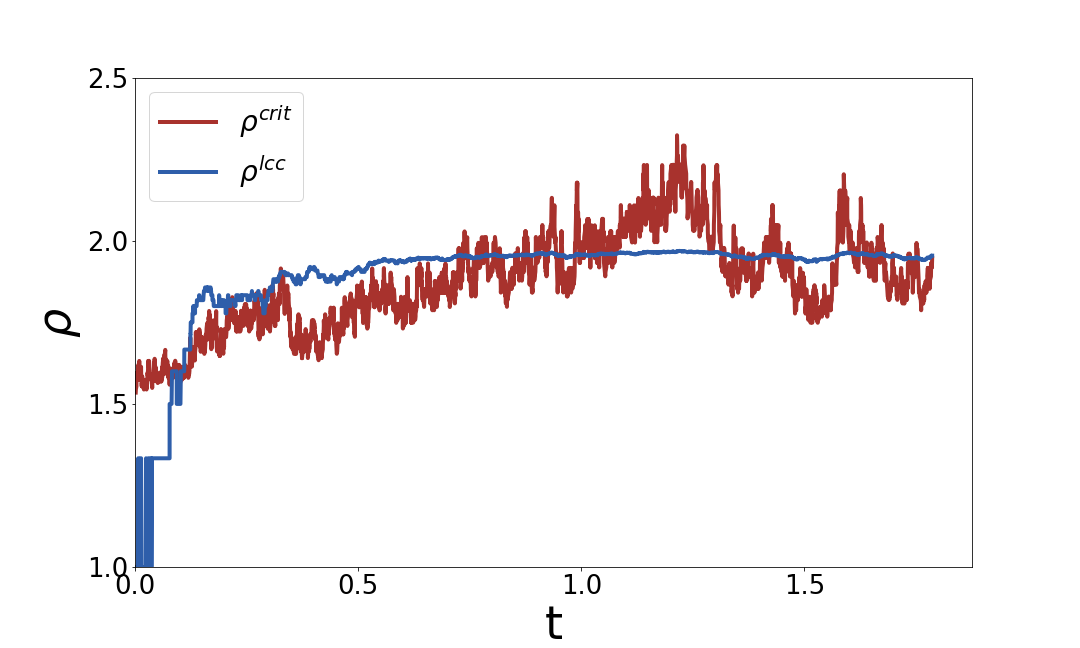}
   \caption{Evolution of the density of the LCC and the critical density in the system over time. $\mu/2$ = 20, $\hat \rho = 21.4$, $\epsilon = 1$, $\gamma = 0.1$, $\beta$ = 0.25, $N = 100$}
  \label{fig:inc-LCC-rho}
\end{figure}

Only groups with $\rho_{k}>\rho_{n}^{\mathrm{cr}}$ are expected to grow incrementally in \emph{size}, while groups with $\rho_{k}<\rho_{n}^{\mathrm{cr}}$ will dissolve. 
On the other hand, $\rho_{k}=2$ for sparse groups. 
Hence, if we expect a stable group, the critical density $\rho_{n}^{\mathrm{cr}}(t)$ should reach approximately 2 for large groups. 
This is indeed shown in Figure~\ref{fig:inc-LCC-rho}, where we plot the density of the largest connected component.
Initially, $\rho_{k}$ is smaller than 2 because of finite size effects.
For instance, for $n_{k}=3$ we would have  $\rho_{k}=2(n_{k}-1)/n_{k}=4/3<2$. 
But after the initial formation of a large group $\rho^{\mathrm{LCC}}$ approaches the value of 2, while $\rho_{n}^{\mathrm{cr}}$ fluctuates around this value.
The fluctuations result from the fact that the largest group continues to have spontaneous additions and dropouts of agents.
We remind that $\hat{\rho}\gg 2$ is compararably large, i.e. for agents the transition rate to leave the group is not negligible.
Because both $n_{0}$ and $\mu_{0}$ are small, changes in these numbers impact the critical density, $\rho^{\mathrm{cr}}_{n}$.
The density of the group, $\rho_{k}$, on the other hand, is not much impacted if a single agent joins or leaves because of the large group size. 

By comparing Figures~\ref{fig:time}, \ref{fig:inc-LCC-rho}, we can estimate the time scales for the dynamics of $n_{k}(t)$ and    $\rho_{k}(t)$.
Obviously, the stable density $\rho_{k}=2$ is reached  much faster, at about $t\approx 0.1$. 
But this still allows the largest group to grow, i.e. $n_{k}(t)$ continues to change until $t\approx 10$.
Therefore, in the following we will analytically describe its dynamics.
If we only consider incremental growth and dissolution processes for groups, Eqs.~\eqref{eq:21-inc}, \eqref{eq:400}, we can write down a dynamics for the expectation value of the group size $\mean{n_{k}}=\sum\nolimits_{k^{\prime}}n_{k^{\prime}}P(n_{k^{\prime}},t)$, by starting from:
\begin{align}
  \label{eq:15}
  \frac{d\mean{n_{k}(t)}}{dt} &= \mean{w^{+}\left[g_{k},g_{1}\right] - w^{-}\left[g_{k},\rho_{k}\right]} \nonumber \\
                              &= \frac{\mean{n_{k}(t)}}{N}\left[ \mean{y_{0}(t)} - e^{
                                \beta\frac{\hat\rho}{\mean{\rho_{k}(t)}}
                                }\right]
\end{align}
If we insert the derived expression for the critical density $\rho_{n}^{\mathrm{cr}}(t)$, Eqn.~\eqref{eq:10}, we can express the dynamcis as:
\begin{align}
  \label{eq:28}
  \frac{d \ln \mean{n_{k}(t)}}{dt} &=  \frac{\beta\hat\rho}{N} \left[ \frac{1}{\mean{\rho_{n}^{\mathrm{cr}}(t)}} - \frac{1}{\mean{\rho_{k}(t)}}  \right]                                
\end{align}
Here we have used a linear expansion of $\exp(x)$ and $\ln(x)$.
Eqn.~\eqref{eq:28} defines a \emph{selection equation} \citep{ebeling2011physics} that couples the growth of all groups via $\rho_{n}^{\mathrm{cr}}$. 
The \emph{size} of group $k$ grows as long as its density is larger than the critical density $\rho_{n}^{\mathrm{cr}}$.
We remind that a similar selection equation appears in nucleation theory, Eqn.~\eqref{eq:32}, where the coupling was given by the critical radius $r_{0}^{\mathrm{cr}}(t)$.

\subsection{Group densification}
\label{sec:group-densification}

We now turn to the second limit case where instead of incremental growth only densification is considered, as shown in Figure~\ref{fig:symbolic representation}(c) and described by Eqn.~\eqref{eq:14}.
We can then calculate the odds ratio as:
\begin{equation}
  \label{eq:700}
  G = \ln \frac{w^{+}\left[g_{k},m_{0}\right]}{w^{-}\left[ g_{k},\rho_{k}\right]}=
  \ln \left[\frac{2m_{0}n_{k}[n_{k}-\rho_{k}]}{N^{2}} \frac{N}{n_{k}\exp\{\beta \hat{\rho}/\rho_{k}\}}\right]
\end{equation}
From $G=0$ we find the critical density for densification: 
\begin{align}
  \label{eq:101ng}
  \rho^{\mathrm{cr}}_{d}(t)=\frac{\beta\hat{\rho}}{\ln \mu_{0}(t) + \ln \omega_{k}(t)}
  \;;\quad \omega_{k}(t)=\left[n_{k}(t)-\rho_{k}(t)\right]
\end{align}
We note that the structure of $\rho_{d}^{\mathrm{cr}}$ is similar to $\rho_{n}^{\mathrm{cr}}$, Eqn.~\eqref{eq:10}. 
$\omega_{k}(t)$ gives the number of ``free'' links \emph{inside}  group $k$, while $n_{0}(t)$ which gives the number of free agents \emph{outside} the group. 
That means, $\omega_{k}(t)$ differs for groups in different configurations ($n_k, m_k$).

How does the additional critical density $\rho_{d}^{\mathrm{cr}}$ influence the growth of groups?
We recall that in the limit of incremental growth only $n_{k}$ changes according to Eqn.~\eqref{eq:28}, while $\rho_{k}$ remains constant. 
In the limit of densification, on the other hand, only the number of links $m_{k}$ inside a group changes, while $n_{k}$ remains constant.
Therefore, we need to develop a dynamics for $m_{k}$ now.

The dynamics for the expectation value of the number of links $\mean{m_{k}}=\sum\nolimits_{k^{\prime}}n_{k^{\prime}}P(m_{k^{\prime}},t)$, follows from:
\begin{align}
  \label{eq:15m}
  \frac{d\mean{m_{k}(t)}}{dt} &= \mean{w^{+}\left[g_{k},m_{0}\right] - w^{-}\left[g_{k},\rho_{k}\right]} \nonumber \\
                              &= \frac{\mean{n_{k}(t)}}{N}\left[ \mean{\mu_{0}(t)\omega_{k}(t)} - e^{
                                \beta\frac{\hat\rho}{\mean{\rho_{k}(t)}}
                                }\right]
\end{align}
If we insert the derived expression for the critical density $\rho_{d}^{\mathrm{cr}}(t)$, Eqn.~\eqref{eq:101ng}, we can express the dynamcis as:
\begin{align}
  \label{eq:28mk0}
  \frac{d \mean{m_{k}(t)}}{dt} &=  \frac{\beta\hat\rho}{N}\mean{n_{k}(t)} \left[ \frac{1}{\mean{\rho_{d}^{\mathrm{cr}}(t)}} - \frac{1}{\mean{\rho_{k}(t)}}  \right]                                
\end{align}
which, in first order approximation $\mean{\rho_{k}}=2\mean{m_{k}}/\mean{n_{k}}$, eventually gives a selection equation for $\mean{\rho_{k}}$:
\begin{align}
  \label{eq:28mk}
  \frac{d \mean{\rho_{k}(t)}}{dt} &=  \frac{\beta\hat\rho}{N} \left[ \frac{1}{\mean{\rho_{d}^{\mathrm{cr}}(t)}} - \frac{1}{\mean{\rho_{k}(t)}}  \right]                                
\end{align}
We note that the two critical densities for incremental growth and for densification are not independent of another, they are related via $\mu_{0}(t)$.
By solving Eqs. \eqref{eq:10}\eqref{eq:101ng} for $\mu_{0}(t)$ we find:
\begin{align}
  \label{eq:37}
  \frac{\beta\hat\rho}{\mean{\rho_{n}^{\mathrm{cr}}(t)}}=
  \frac{\beta\hat\rho}{\mean{\rho_{d}^{\mathrm{cr}}(t)}} -\ln{\frac{\omega_{k}}{n_{0}}}
\end{align}
After the initial formation of a larger number of groups,
agents will leave groups with subcritical density, to join  larger groups with supercritical density.
This increases the size of supercritical groups at the expense of subcritical groups which dissolve eventually.
This reduces the number of groups and will be described analytically  in the following.

\subsection{Competition between groups}
\label{sec:ostwald-ripening}

\begin{figure}[htbp]   \centering
   \includegraphics[width=0.45\textwidth]{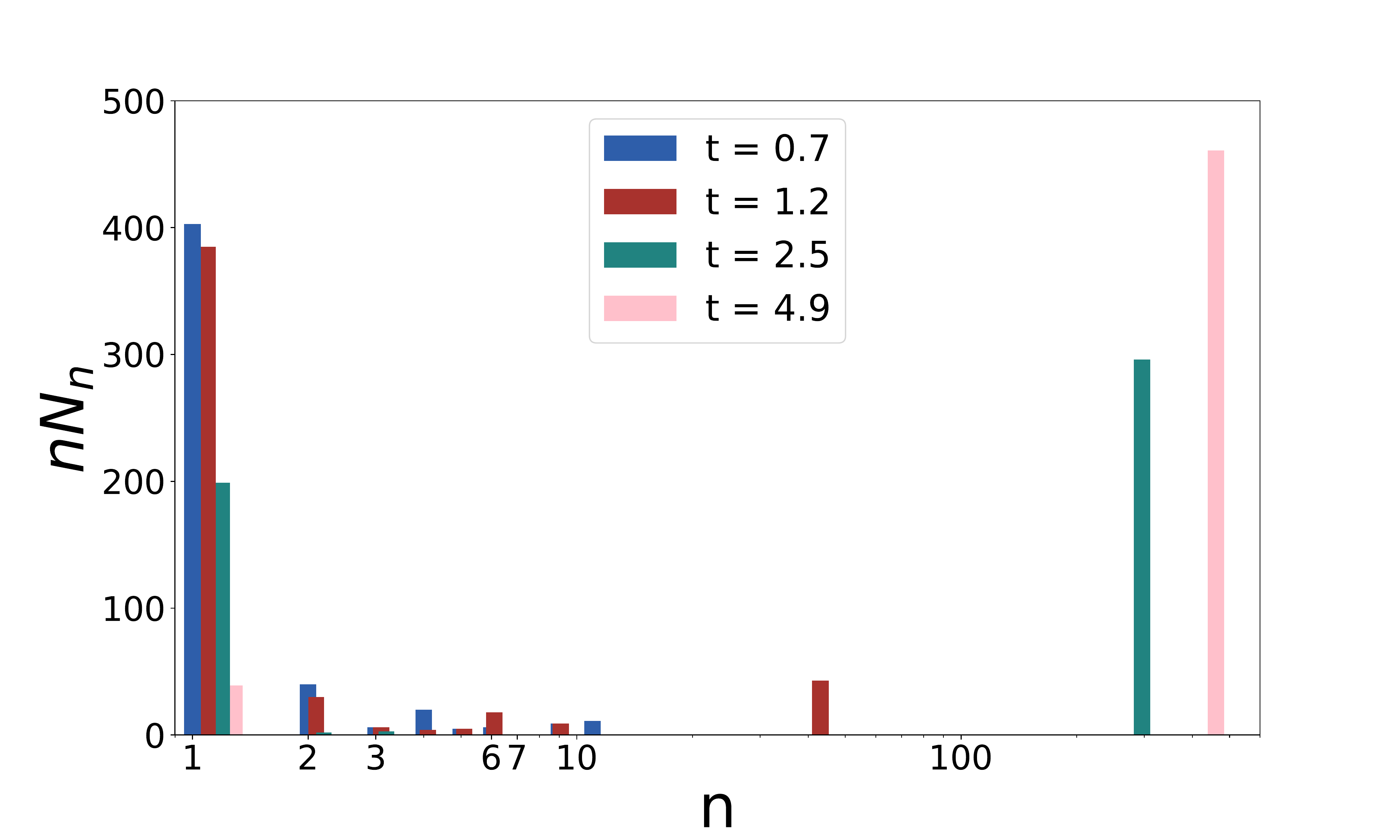}(a)
   \includegraphics[width=0.45\textwidth]{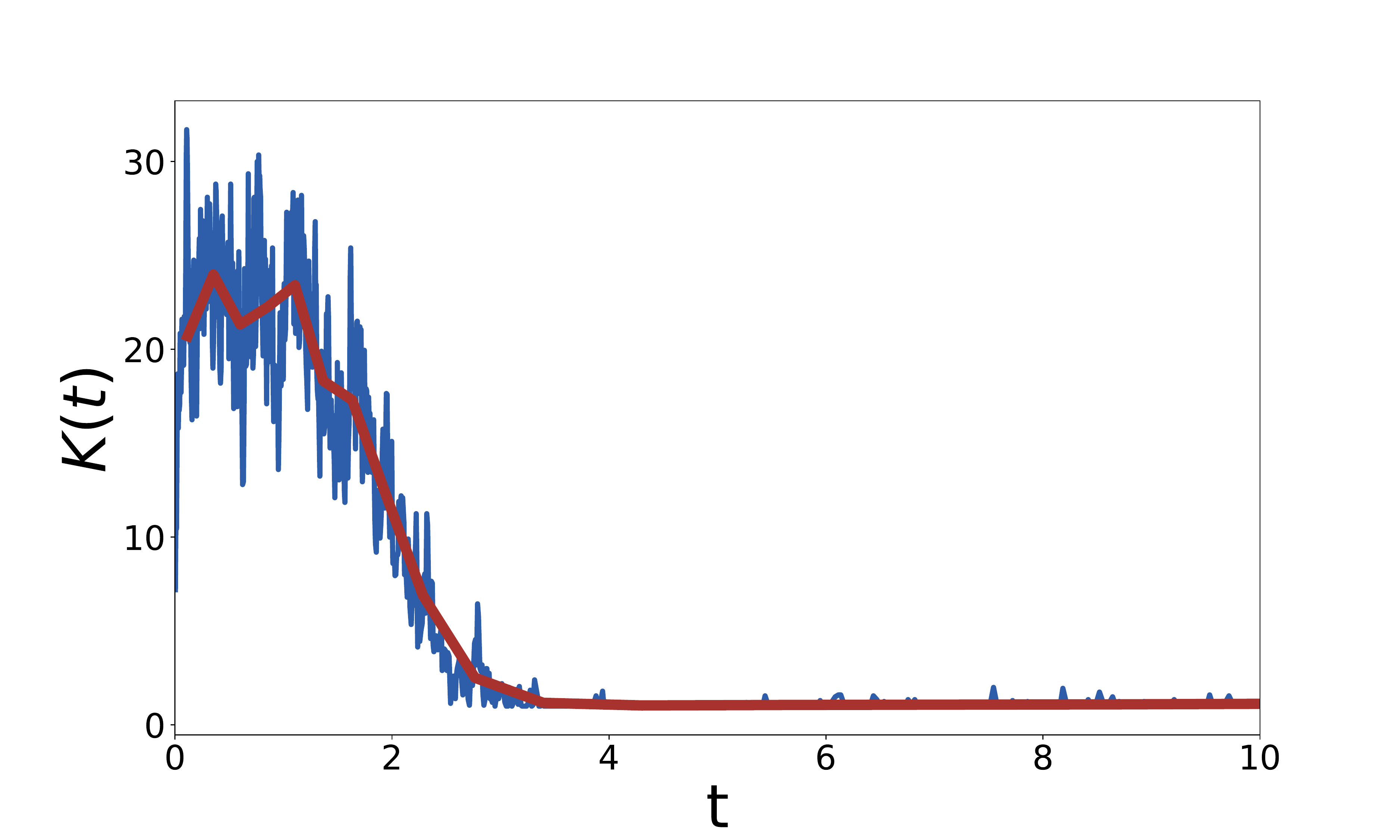}(b)
   \caption{(a) Distribution of agents in $N_{n}$ groups of size $n$ (in $\log$ scale) at various time steps.
     (b) Total number of groups, $K$, over time.
     Parameters: $N$ = 500, $\mu/2$ = 5, $\hat \rho$ = 17.2, $\epsilon$ = 1, $\gamma$ = 0.1, $\alpha$ = 0.9, $\beta$ = 0.4.
   }
  \label{fig:group-dist}
\end{figure}

Figure~\ref{fig:group-dist}(a) shows the evolution of the group distribution for $N=500$, by plotting the quantity $n_{k}N_{n_{k}}$ over time.
$n_{k}$ is the size of group $k$ and $N_{n_{k}}$ is the number of groups with size $n_{k}$.
In an early stage, we find a large number of smaller groups of subcritical size, $n_{k}<6$.
Because $N$ is comparably small, only 
few of these groups reach a supercritical size.
Only one continues to grow incrementally and via coagulation, while the subcritical groups either dissolve gradually or fragment. 
For larger systems, we find more supercritical groups, but this does not affect our principal discussion.

Figure~\ref{fig:group-dist}(b) shows the total number of groups $K$ over time. 
Starting from zero, in a very short time, $ t\leq 2$, up to 30 groups of small size are formed
In a second phase, $2 \leq t \leq 4$, this number decreases. 
In the end, we find one large group surrounded by a small number of free agents, as illustrated in Figure~\ref{fig:nucleation}(c). 

To describe this evolution, we need to consider the dynamics for $n_{k}(t)$, Eqn.~\eqref{eq:28}.
Densification does not change the size or the number of groups, but it can stabilize established groups. 
The smallest group size is $n_{k}=2$, otherwise agents are considered as free agents and counted in $n_{0}$.
Hence if $n_{k}\to 1$, the group has dissolved and $K(t)$ is diminished by 1.
According to Eqn.~\eqref{eq:28}, $n_{k}$ decreases if $\rho_{k}<\rho_{n}^{\mathrm{cr}}$.
This does not exclude processes to increase $\rho_{k}$ via densification, therefore we base our further discussion on the selection Eqn.~\eqref{eq:28} for $n_{k}(t)$. 

We start from the conservation of agents, Eqn.~\eqref{eq:1}, which takes the sum over all existing groups:
$\sum\nolimits_{k}n_{k}=[N-n_{0}]$. 
This yields ${d n_{0}(t)}/{dt}= - \sum\nolimits_{k} {dn_{k}}/{dt}$. 
Let us define the average group size $\bar{n}$ and the average group density $\bar{\rho}$ as:
\begin{align}
  \label{eq:36}
  \bar{n}(t)= \frac{1}{K(t)}{\sum\nolimits_{k} n_{k}(t)}=\frac{1}{K(t)}[N-n_{0}]\;;\quad
  \bar{\rho}(t)= \frac{1}{K(t)}{\sum\nolimits_{k} \rho_{k}(t)}
\end{align}
where $K(t)$ is the total number of groups at a given time.
Further, we assume that in a first order approximation $\sum\nolimits_{k}{n_{k}}/{\rho_{k}}=K \bar{n}/\bar{\rho}$. 
Then we obtain from Eqn.~\eqref{eq:28}: 
\begin{align}
  \label{eq:12}
- \frac{dn_{0}(t)}{dt}= & \sum\nolimits_{k} \frac{\beta \hat\rho}{N}n_{k}(t)\left[\frac{1}{\rho_{n}^{\mathrm{cr}}(t)}-\frac{1}{\rho_{k}(t)}\right] =   \frac{\beta \hat\rho}{N} K(t) \bar{n}(t)  \frac{1}{\rho_{n}^{\mathrm{cr}}(t)}
      - \frac{\beta \hat\rho}{N} K(t) \frac{\bar{n}(t)}{\bar{\rho}(t)}
\end{align}
which can be solved for $\rho_{n}^{\mathrm{cr}}(t)$:
\begin{align}
  \label{eq:29}
  \frac{1}{\rho_{n}^{\mathrm{cr}}(t)}=
  \frac{1}{\bar\rho(t)}
  - \frac{N}{\beta\hat \rho}\frac{1}{K(t)\bar{n}(t)}\frac{d n_{0}(t)}{dt}
\end{align}
This allows us to replace $1/\rho_{n}^{\mathrm{cr}}$ in Eqn. \eqref{eq:28} such that we finally arrive at:
\begin{align}
  \label{eq:25}
  \frac{d \ln n_{k}(t)}{dt}= \frac{\beta\hat\rho}{N}\left[\frac{1}{\bar \rho(t)}-\frac{1}{\rho_{k}(t)}\right] -\frac{1}{K(t)\bar{n}(t)}\frac{d n_{0}}{dt}
\end{align}
Eqn.~\eqref{eq:25} has several advantages:
(i) Instead of the well defined, but unknown, variable $\rho_{n}^{\mathrm{cr}}$, which characterizes the system via $y_{0}(t)$, we can use the average group density $\bar{\rho}$, which characterizes the groups.
Note that $\bar{\rho}$ represents all groups with subcritical and supercritical densities and  also reflects any densification of groups, i.e. it is the appropriate aggregated variable.

(ii)  For the dynamics of Eqn.~\eqref{eq:25} we can distinguish two stages of the phase transition.
In the first stage of \emph{group formation}, groups are small and of comparable density.
Hence $\rho_{k}\approx \bar{\rho}$ and the first term vanishes. 
Therefore the dynamics is dominated by the last term:   
 $(1/K\bar{n})({d n_{0}}/{dt})$ decreases rapidly because the number of free agents, $n_{0}$, decreases fast \emph{and} the number of groups, $K$, as well as the average group size, $\bar{n}$, grows.
 In the second phase of \emph{group competition} we can already assume that ${d n_{0}}/{dt}\approx 0$.
 Hence the dynamics is dominated by the first term in square brackets, which describe a \emph{slow} selection process.  

(iii) The meaning of competition and selection is made explicit in the dynamics:
The growth of groups is coupled by the mean density $\bar{\rho}$. 
Because of ${d n_{0}}/{dt}\approx 0$,  the growth of groups with a larger density can only occur because of the dissolution of groups with a smaller density.
This process continues as long as $\bar{\rho}$ can still grow.
The dynamics reaches a (theoretical) stationary state if only one group is left which defines the average density, $\bar\rho=\rho_{k}$.
This assumes that all agents can form links, which implies $\epsilon=1$.
If the tolerance threshold $\epsilon$ is low, smaller groups will still dissolve.
But the free agents cannot join the larger group because of the differences in opinions.
Therefore they remain as isolated agents, as shown in Figure~\ref{fig:nucleation}(c).

\section{Discussion}
\label{sec:discussion-1}

\subsection{A coherent view on the dynamics of group formation}
\label{sec:coher-view-dynam}

We have now all pieces together to present a coherent view of the group formation and competition, with particular emphasis on the nucleation scenario.
We structure the discussion with respect to the various critical densities that are introduced in this paper. 
The first one is $\hat\rho=2c/b$, which is constant and defines the minimum density for a group to have a \emph{non-negative utility}.
Social agents evaluate costs $c$ and benefits of being part of a group $b$ and will leave if they do not gain a utility from this, Eqn.~\eqref{eq:400}.

\paragraph{$\hat\rho<\rho_{k}$.}
Groups need to have a density $\rho_{k}>\hat\rho$ to remain or even grow, otherwise they will dissolve.
For the percolation scenario $\hat\rho=0$, i.e. groups will never dissolve.
But it needs a critical supersaturation $y>1$, Eqn.~\eqref{eq:35}, to obtain a large group that connects almost all agents.
Yet, because $y$ is comparably small, the largest connected group is still very sparse, as Figure~\ref{fig:perc}(c) shows.
For the scenario of spinodal decomposition, the condition  $\rho_{k}>\hat\rho$ is also fulfilled, because $\hat \rho$ is very small.
But, different from percolation, the initial saturation $y$ is very large.
Thus, we see the fast formation of one large compact group.

\begin{figure}[htbp] \centering
 \includegraphics[width=0.45\textwidth]{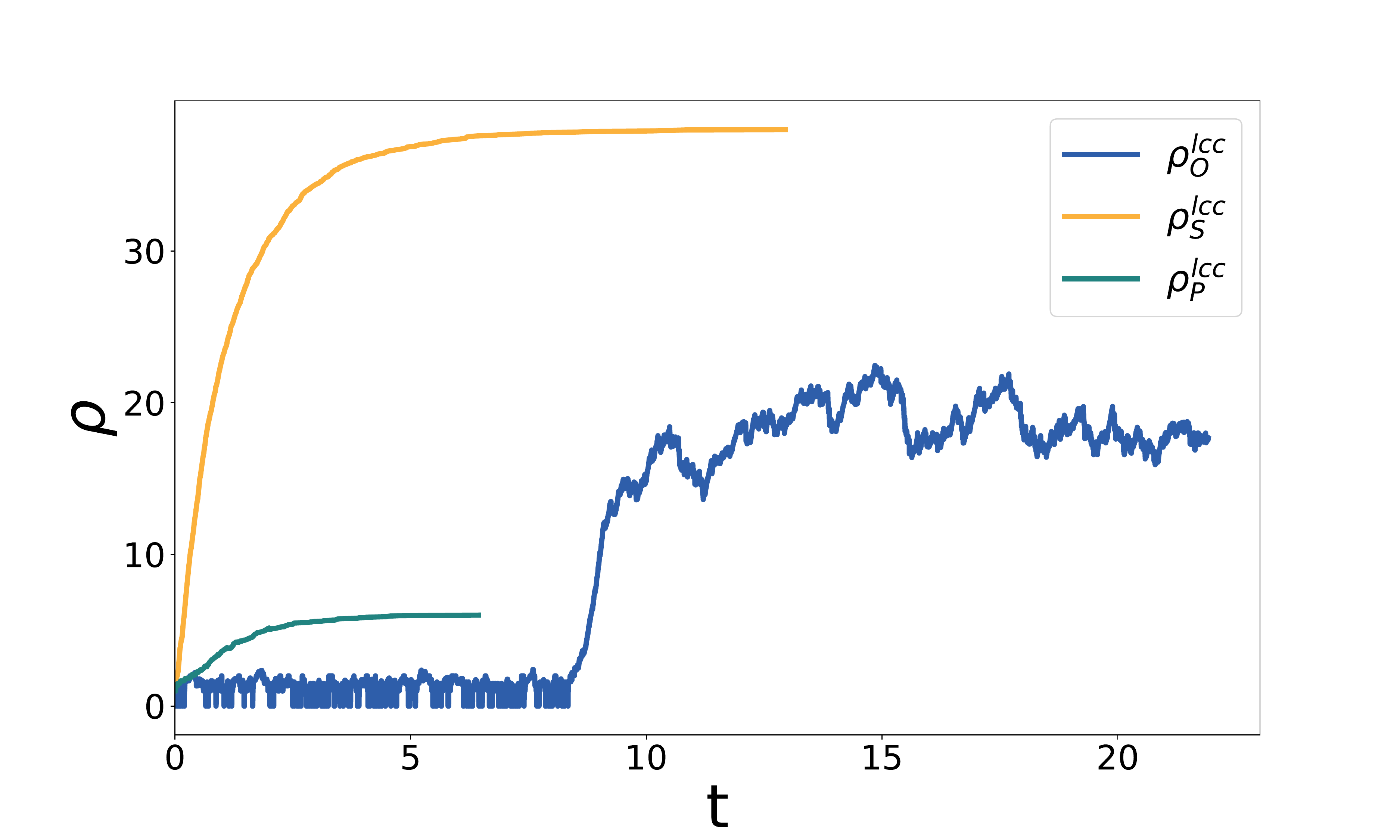}
 \caption{Density of the largest connected group over time for percolation, $\rho^{LCC}_{P}$ ($\mu_P$/2 = 3, $c_{P}$ = 0), spinodal decomposition, $\rho^{LCC}_{S}$ ($\mu_S$/2 = 19, $c_{S}$ = 0) and nucleation, $\rho^{LCC}_{O}$ ($\mu_O/$2 = 19, $\hat \rho$ = 17.2). Parameters: $\epsilon$ = 1, $\gamma$ = 0.1, $\beta$ = 0.4.
 }
  \label{fig:qt}
\end{figure}

The differences can be also observed in Figure~\ref{fig:qt} where we plot the density of the largest group over time.
Percolation occurs rather fast, but the largest group reaches only a small final density because it is a sparse network.
Spinodal decomposition occurs very fast because of the large initial instability.
Because the initial supersaturation is high, i.e. a large number of links is potentially available, the final density of the largest group is also high. 

\begin{figure}[htbp] \centering
   \includegraphics[width=0.45\textwidth]{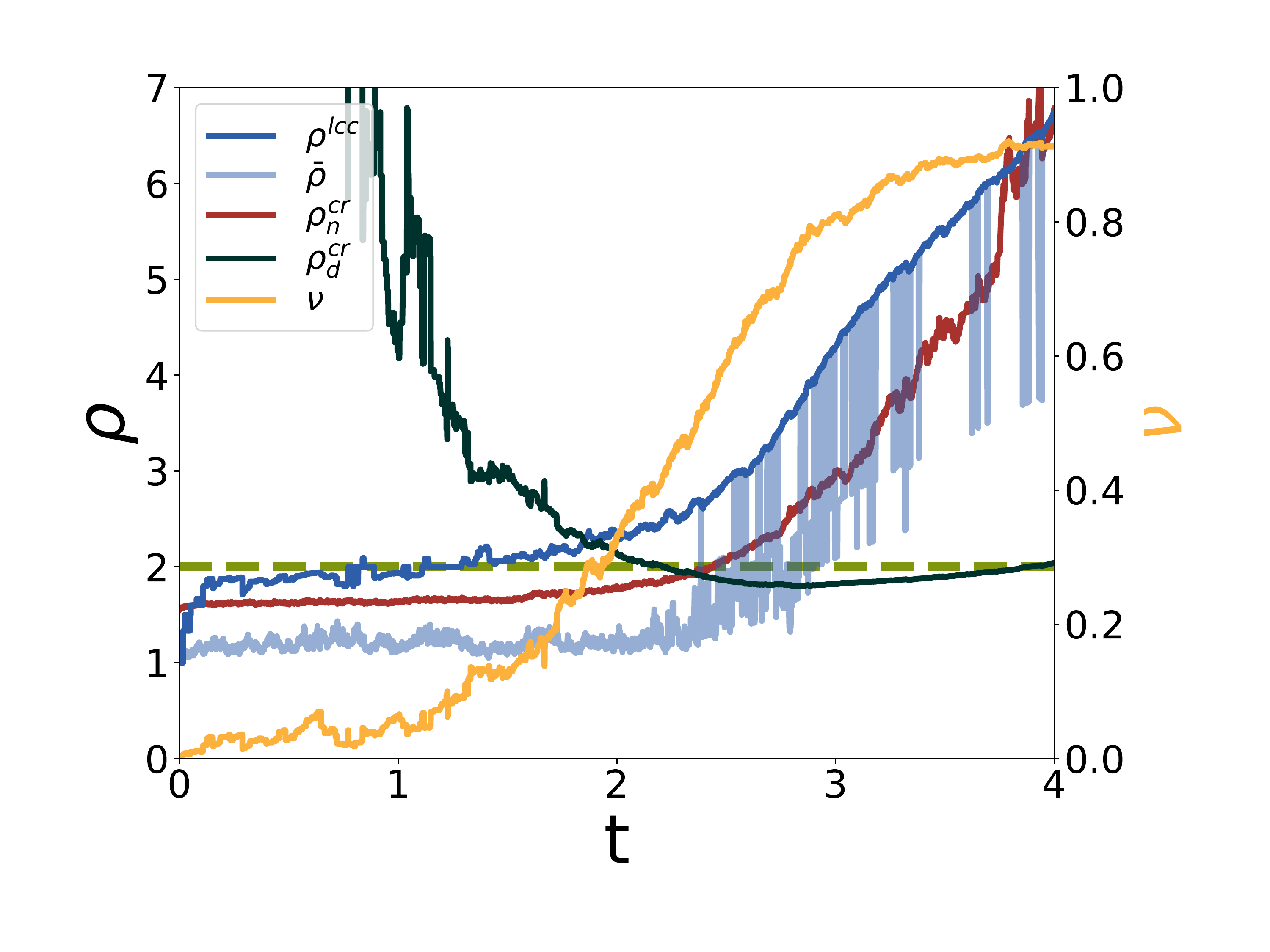}(a)
   \includegraphics[width=0.45\textwidth]{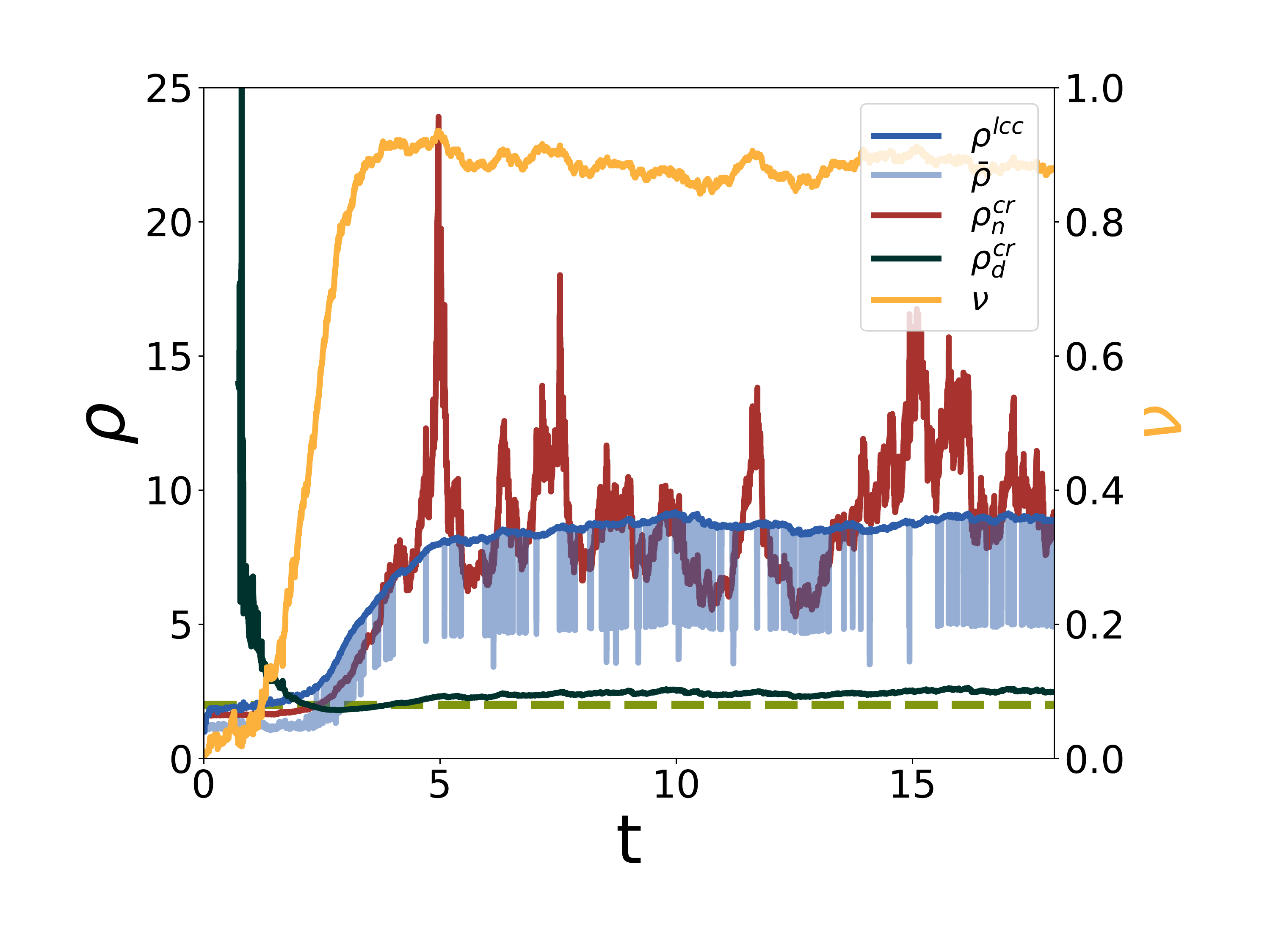}(b)
   \caption{Evolution of the density, $\rho^{\mathrm{LCC}}$, and the fraction, $\nu$, of the largest connected group, the mean density, $\bar\rho$, the critical densities, $\rho_{n}^{\mathrm{cr}}$, $\rho_{d}^{\mathrm{cr}}$. (a) Early time period, (b) long term dynamics. 
     Parameters: $N$ = 500, $\mu/2$ = 5, $\hat \rho$ = 17.2, $\epsilon$ = 1, $\gamma$ = 0.1, $\alpha$ = 0.9, $\beta$ = 0.4.
   }
  \label{fig:500Agents-new}
\end{figure}

\paragraph{$\rho_{k}<\hat\rho$.}
For the nucleation scenario, the condition  $\rho_{k}>\hat\rho$ is \emph{never} fulfilled.
That means, different from the other two scenarios, we will \emph{always} observe that agents leave a group.
Therefore, even the largest group experiences fluctuations in size.
As another consequence, groups can form initially only by means of fluctuations and have to overcome an energy barrier, quantified  by the critical density $\rho_{n}^{\mathrm{cr}}$, Eqn.~\eqref{eq:10} which 
 is plotted in Figure~\ref{fig:500Agents-new}.

Hence, in this scenario it takes much longer before a large connected group can establish, which is also shown in Figure~\ref{fig:qt}.
The final density of the group is largely determined by the initial supersaturation, i.e. the number of potentially available links per agent.
The large group is not as sparse as in the case of percolation, but not as dense as in the case of spinodal decomposition. 

\paragraph{$\rho_{n}^{\mathrm{cr}}(t) <\rho_{k}< \rho_{d}^{\mathrm{cr}}(t)$.}

If  $\rho_{k}<\hat\rho$, the two critical densities for growth, $\rho_{n}^{\mathrm{cr}}$, Eqn.~\eqref{eq:10}, and for densification, $\rho_{d}^{\mathrm{cr}}$, Eqn.~\eqref{eq:101ng}, come into play.
These are plotted in Figure~\ref{fig:500Agents-new}. 

As we show in Figure \ref{fig:500Agents-new}(a), during the initial stage of group formation, $t\leq 2$, $\rho_{d}^{\mathrm{cr}}(t)$ is very large.
This implies that initially densification \emph{cannot} take place because $\rho_{k}(t)$ is too small.
As a consequence, the dominating process in the initial stage is the incremental growth of $n_{k}(t)$.
We see this in Figure \ref{fig:500Agents-new}(a), where relative size of the largest group, $\nu=n^{\mathrm{LCC}}/N$ continuously increases, while its \emph{density} has the constant value, $\rho^{\mathrm{LCC}}=2$, for incremental growth.  
Also the critical density $\rho_{n}^{\mathrm{cr}}$ is rather small, below 2, which allows the largest group to grow incrementally, according to the selection Eqn.~\eqref{eq:28}.

\paragraph{$\rho_{d}^{\mathrm{cr}}(t) < \rho_{n}^{\mathrm{cr}}(t)<\rho_{k}$.}
The dynamics changes from incremental growth to densification once $\rho_{d}^{\mathrm{cr}}(t)$ becomes small enough because the link supersaturation $\mu_{0}(t)$ has decreased in the course of group formation.
Precisely, once $\rho_{d}^{\mathrm{cr}}\leq 2$, the process of densification starts to dominate .
This does not mean that no group grows further.
In fact, the largest group now grows \emph{both} in size and in density, as shown in Figure \ref{fig:500Agents-new}(a) because it fulfills the condition $\rho_{d}^{\mathrm{cr}}(t) < \rho_{d}^{\mathrm{cr}}(t)<\rho_{k}$.

\paragraph{$\rho_{d}^{\mathrm{cr}}(t) <\rho_{k}< \rho_{n}^{\mathrm{cr}}(t)$.}

The growth and densification of the largest group decreases the supersaturation $y_{0}=\mu_{0}n_{0}$, hence $\rho^{\mathrm{cr}}_{n}$ increases considerably.
This then stops the further \emph{growth} of groups with $\rho_{k}< \rho_{n}^{\mathrm{cr}}(t)$.
This does not exclude that these groups can densify, because $\rho_{d}^{\mathrm{cr}}(t) <\rho_{k}$ holds.
But the group \emph{size} still diminishes because $\rho_{k}<\rho_{n}^{\mathrm{cr}}$,

\paragraph{$\bar{\rho}(t)\approx\rho_{k}$. }

The third relevant density is the average density $\bar{\rho}(t)$, Eqn.~\eqref{eq:36}.
As an average over groups, $\bar{\rho}(t)$  describes the distribution of groups at a given time.
In Figure~\ref{fig:500Agents-new}(a) we observe that, for $t\leq 2$,  $\bar{\rho}$ is almost constant, has a small value and fluctuates very little.
This means that groups are small and of similar size and have not yet reached the density for sparse, but large clusters, $\rho_{k}=2$. 
This observation further justifies why we have dropped, for the first stage, the first term of the competition Eqn.~\eqref{eq:25}, because $\rho_{k}\approx \bar{\rho}$.

\paragraph{$\bar{\rho}(t)<\rho^{\mathrm{LCC}}$. }
For the second stage, $2<t<4$, we observe instead a continuous growth of $\bar\rho(t)$ over time. 
This is mainly dominated by the density of the largest group, but fluctuates because small groups disappear.
We further observe that for the largest group always $\bar{\rho}(t)<\rho^{\mathrm{LCC}}$ holds.
That means, this group can grow further.
Groups with $\rho_{k}<\bar{\rho}(t)$, on the other hand, cannot grow but will dissolve.
This is reflected in the first term of the competition Eqn.~\eqref{eq:25}, which now plays the mayor role in the dynamics.

\paragraph{$\rho^{\mathrm{cr}}_{n}(t)$ vs. $\rho^{\mathrm{cr}}_{d}(t)$. }
For the formation of new groups the relevant control parameter is $\rho^{\mathrm{cr}}_{n}(t)$.
It determines the critical density that new groups need to reach if they want to establish. 
Whereas $\rho^{\mathrm{cr}}_{d}(t)$ describes the evolution of the \emph{group}, 
$\rho^{\mathrm{cr}}_{n}(t)$ describes the evolution of the \emph{system} because it depends on the supersaturation $y_{0}(t)$.
This way, $\rho_{n}^{\mathrm{cr}}(t)$ couples the growth of all groups via Eqn.~\eqref{eq:28}.  
Because $y_{0}(t)$ continuously decreases during the phase transition,  $\rho_{n}^{\mathrm{cr}}(t)$ increases over time and sets a limit to the spontaneous group formation at a later time $t$.

$\rho_{d}^{\mathrm{cr}}$, on the other hand, is relevant only for the internal stability of the group during the second stage. 
During the first stage densification does not play a role because 
$\rho_{k}<\rho_{d}^{\mathrm{cr}}$.
During the second period densification becomes important, but only to some degree, because  
the group \emph{size} still diminishes as long as $\rho_{d}^{\mathrm{cr}}<\rho_{k}<\rho_{n}^{\mathrm{cr}}$.

Figure~\ref{fig:500Agents-new}(b) shows the long-term dynamics of the process described.
It allows to clearly separate the two stages: a \emph{short initial period}, $t\leq 2$ in which the formation of sparse groups dominates.
This is followed by a \emph{second stage} of about the same duration $2\leq t\leq 4$, which leads to the establishment of a large group.
The precise values for the duration of course depends on the system size, for which we have chosen $N=500$ instead of $N=100$ from before. 
During this second stage the number of groups reduces and the distribution of groups changes as plotted in Figure~\ref{fig:group-dist}.

\subsection{The structure of social groups}
\label{sec:different}

The three different scenarios,
percolation, spinodal decomposition and nucleation, determine the network structure of groups, as detailed in the parameter plot of Figure~\ref{fig: phaseDiagHighTension}(right).
The above discussion has assumed that all agents can interact and form links unconditionally. 
But groups are also structured by the underlying social process of link formation, which we described by means of opinions, $x_{i}(t)$.
These are internal variables of agents that determine whether two agents can interact, which in turn is the precondition of link formation and  group membership.
These opinions are not only changed in bilateral interactions, they are further influenced by the group.
That lead to the concept of an effective opinion that weights individual opinions against the average opinion of a group, using an additional parameter $\alpha$ for group influence.

As we have demonstrated in Figure~\ref{fig: phaseDiagHighTension}(left), restrictions in the interactions of agents, expressed by the tolerance threshold $\epsilon$, can lead to very different macroscopic patterns.
Instead of one large connected group, we very often find that several  groups with 
different opinions finally coexist if $\epsilon$ is small.
In this case, the selection Eqs. \eqref{eq:28}, \eqref{eq:25} no longer hold.
At least for the percolation case we are able to derive a relation, Eqn.~\eqref{eq:3}, to determine whether a given initial link density $\mu$ and a given tolerance threshold $\epsilon$ would still allow for the formation of a large connected group. 

If we take the social perspective then, different from physics, the final equilibrium state is less interesting than the dynamics that potentially lead to it.
Social systems are non-equilibrium systems that adapt and evolve before an equilibrium is reached.
Our paper therefore investigates the \emph{principal} ability of agents to form one large group, or to coexist in several separated groups.
A state where all agents are in the same group is rather unrealistic from a social perspective.
In this respect, the nucleation scenario is the most promising one, as it leads to a distribution of groups that form spontaneously and compete for agents to grow.

The nucleation scenario also has the advantage to consider agents \emph{leaving} the group if they experience a negative group utility.
Such model features can bridge between our rather abstract approach and social processes of group formation.
Agents are allowed to constantly reevaluate their belonging to the social group.
We could further consider that agents reevaluate their relations to those agents they are linked to.
They could then delete a link if the difference between their current opinion and the opinions of their group members has reached a critical threshold \citep{groeber_2014}.
Also more complex decision rules to establish or to delete a link can be considered \citep{Battiston2009,Centola2007}. 

Another model feature with relevance to social systems is the explicit consideration of a \emph{finite} number of agents and links.
Thermodynamic models of phase transitions usually assume the limit of infinite systems.
We instead address that social processes build on \emph{limited} resources, be it available individuals or the ability to maintain social relations.
Therefore, we decided on purpose to present simulations with 100 or 500 agents. 
The depletion of these resources couples the dynamics of different groups on the systemic level, albeit in an indirect manner.
It also decreases the chances for new groups to form at later stages.
Hence, our model reflects the \emph{first mover advantage}: groups that form early have a larger chance to reach a supercritical size or density.

Our model of social group formation also considers that groups can \emph{merge} or \emph{split} into fractions.
The merger process, which is called coagulation in a physical context, is an effective mechanism to overcome the critical group size because it increases the group size not incrementally, but in larger steps.
In our model the possibility for mergers is principally restricted by the opinion dynamics.
If we consider a strong group influence, all opinions inside a group quickly converge to their group average.
If the tolerance threshold is low, groups with very different opinions can no longer merge, so they remain in coexistence, as nicely shown in Figure~\ref{fig:thres-t}(c).
This has indeed analogies to social or economic systems where \emph{local cultures} \citep{groeber_2009_2} inside groups or firms impact the success of mergers and acquisitions. 

Fragmentation is not a separate process but follows from the dissolution of groups, Eqn.~\eqref{eq:400}.
If a link is removed, with a small probability the group breaks into pieces.
This is more likely if groups have a small density.
Sparse groups appear if two conditions are fulfilled: a rather low link supersaturation $\mu$ close to the percolation threshold and a rather large $\epsilon$, i.e. all links are accepted.
For small $\epsilon$, we obtain more compact groups because links within the group are more likely than between groups. 

Eventually, we emphasize that one of the key variables of our model, the group \emph{density} $\rho_{k}$, is particularly relevant in a social context.
In physical systems droplets are described as compact, homogeneous, spherical clusters with a radius $r_{k}$ and a fixed density $\rho_{\alpha}$, which is the same for all droplets.
Social groups, on the other hand, are neither compact, nor spherical, nor homogeneous.
They are like small social networks, more precisely disconnected components of a large social network.
Therefore they need to be characterized by two variables, the number of agents, $n_{k}$, and the number of links, $m_{k}$, which together define a time dependent group density.
$\rho_{k}(t)$ in fact determines the \emph{quality}, or the fitness, of a social group, which depends on the \emph{relations} between group members rather than on the sheer number of members.
Given a fixed group size, the number of realized relations makes all the difference for exchanging information, collaborating and sharing resources.
This has been accounted for in our model by defining a group utility based on the group density.

While being relevant in a social context, our model allows for analytic investigations to relate  the group density $\rho_{k}$ to different critical densities, $\hat{\rho}$, $\rho_{n}^{\mathrm{cr}}(t)$, $\rho_{n}^{\mathrm{cr}}(t)$ and $\bar\rho(t)$ which compress important information about the dynamics of social groups.
This way, we could compactly describe the collective dynamics of all groups in a selection Eqn.~\eqref{eq:25} that allows to separate early and late influences.
In particular, it formally describes the \emph{winner takes all} dynamics that is known from many social and economic systems \citep{ebeling2011physics,Schweitzer2020}.

\subsection{Outlook}
\label{sec:outlook}

As a main conceptual contribution, in our paper we formally introduce  the \emph{feedback} between the formation of social groups and the opinion dynamics of individuals.
According to the principle of homophily agents interact more if they are more similar with respect to some features.
We capture these features in a rather abstract notion of ``opinion''.
Agents will establish social relations, i.e. links, if their similarity in opinions allows them to interact.
Groups are formed based on these social relations.
Once they are established, they continue to influence the opinion of agents, this way impacting the possibility of agents to form new links.
Therefore, in our model we observe the \emph{coevolution} between social group structures and opinions.
Different from simple models of network formation, the probability that two agents form a link is no longer a global and constant parameter, but (i) depends on the agents, and (ii) evolves over time.

In this paper, we have used only a simplified characterization of agents by means of a continuous scalar variable, $x_{i}(t)$.
In a next step, we will extend our model by considering \emph{multidimensional} opinions \citep{schweighofer_2020}.
This is not just an upscaling of the current model, but will in fact  allow for more diverse group configurations and for new system states.
The multidimensional representation reflects the opinion of a single agent with respect to different issues.
Two agents can have similar opinions about one particular issue, but vastly deviating ones about another issue.
This then leads to the interesting question how agents settle their relations in such cases.
Here psychological concepts like dissonance minimization \citep{festinger_1957,groeber_2014} or 
structural balance theory \citep{heider1946attitudes, flache2017,schweighofer_2020_2,gorski2020homophily,Agbanusi2018,antal3, Kulakowski2019} come into play.
With respect to social group formation, such conflicting situations would allow  agents to form a social group with a focus only on the one issue they agree on, for example a movement for environmental protection.
But regarding another issue, agents would join a different group.

To model these more complex situations, we will explore a multilayer network representation \citep{hu2011percolation,murase2014multilayer,Kivela2014,Scholtes2016,
  Atkisson2019,
  Gorski2017}.
The nodes in each layer, i.e. the agents, are identical, but their relations in each layer can be different because 
each layer contains information about one particular issue only.
Decoupled layers would display a dynamics of group formation similar to the one discussed in this paper.
However, in addition to the \emph{intralayer} dynamics, we have to consider the \emph{interlayer} dynamics of coupled layers.
That is, how do the social dynamics of groups and opinions on one layer impact the dynamics on other layers?
Combining multilayer network models with models of group formation and opinion  dynamics will allow us to address long standing questions about \emph{issue alignment} \citep{schweighofer_2020}, i.e. how opinions on different issues influence another, on the emergence of social movements, or on conflict resolution and polarization of opinions in a novel manner.

\subsection*{Acknowledgements}
\label{sec:acknowledgements}

F.S. gratefully acknowledges early discussions on the dynamics of phase transitions in finite systems with four colleagues, who celebrate their milestone birthday in 2021: Heinz Ulbricht (90), Werner Ebeling (85), Gerd Röpke (80), Jürn Schmelzer (70). Their scientific advice has inspired the current manuscript. 

\begin{appendix}
  
\section*{Appendix A}

Table \ref{tab: summary parameters} below summarizes the correspondence between the mechanisms incorporated in our model, their rationale and their social relevance:

 \renewcommand{\arraystretch}{2}
\begin{table}[httb]
\begin{center}\footnotesize
  \begin{tabular}{@{} p{2.4cm}  p{3cm}  p{3cm}  p{3cm}  p{3cm} @{}} \toprule
 & Opinion formation & Group influence & Group formation & Network formation  \\[5pt]
  \midrule
Social relevance & \makecell[l]{- social influence ($\gamma$) \\ - tolerance ($\epsilon$) \\
    }&
      \makecell[l]{- local cultures ($\alpha$) \\ - social norms ($\alpha$)
        }&
       \makecell[l]{- communication ($b$) \\
    - maintain group ($c$)} &  \makecell[l]{- randomness ($\beta$) \\
    - Dunbar's \# ($M$)
    }\\[5pt]
Rationale & \makecell[l]{- restrict interaction
      \\
    - consensus } &
        \makecell[l]{- group coherence \\ - group relevance}& \makecell[l]{- cost/benefit \\ - group utility}&
     \makecell[l]{- unpredictability\\
    - finite size}   \\[5pt]

    Parameters & $\gamma$  ,  $\epsilon$ & $\alpha$ & $(c,b)$ or $\hat\rho$ & $\beta$  ,  $(N,M)$\\[5pt]
    \bottomrule
  \end{tabular}
\end{center}
\caption{Summary of model components}
\label{tab: summary parameters}
\end{table}
\section*{Appendix B}

In the following we summarize the master equation formalism for considering the group distribution.
The discussion is restricted to 
the case of $\epsilon = 1$ and incremental growth and dissolution as the only possible transitions.

This perspective is based on $P[\pmb{N},t]$, the probability to find the system at time $t$ with a group configuration $\pmb{N}$.
Here, $\pmb{N}=[N_{1,0},N_{2,1},N_{3,2},N_{3,3},...,N_{N,M}]$ is a vector with elements $N_{n,m}$ counting the number of groups of size $n$ with $m$ links.
With this, the total number of links and agents,  Eqn.~\eqref{eq:1}, can be rewritten as: 
\begin{align}
  \label{eq:A1}
  M=m_{0}(t) +\sum_{n,m}N_{n,m}(t)m\;;\quad
   N=n_{0}(t) +\sum_{n,m}N_{n,m}(t)n
\end{align}
where $n_{0}(t)=N_{1,0}(t)$ is the number of free agents as defined in the main text.
The transition rates for growth can then be written as:
\begin{align}
  \label{A3}
w^{+}_{n,m}[N_{1,0} N_{n,m}] \propto \frac{2 m_{0}}{N^{2}} (nN_{n,m})n_{0}
\end{align}
and the transition rates for spontaneous leaving  read: 
\begin{align}
  \label{eq:A4}
w^{-}_{n,m}[N_{n,m}] \propto \frac{nN_{n,m}}{N}  \exp{\left\{\beta\frac{\hat{\rho}}{\rho}\right\}}
\end{align}
Using the above defined transition rates, we can write down the master equation for changes in the group distribution:
\begin{align}
  \label{eq:8}
  \frac{\partial P(\pmb{N},t)}{\partial t}=& \sum\limits_{\pmb{N^{\prime}}\neq \pmb{N}} w[\pmb{N}|\pmb{N^{\prime}}]P(\pmb{N^{\prime}},t) -
                                            w[\pmb{N^{\prime}}|\pmb{N}]P(\pmb{N},t)
\end{align}
which in explicit form can be written as follows:
\begin{multline}
  \label{eq:8}
  \frac{\partial P(N_{1,0},N_{2,1},N_{3,2},N_{3,3}...,t)}{\partial t}=  w^{-}_{2,1}[N_{2,1}+1]P(N_{1,0}-2,N_{2,1}+1...,t) - \nonumber \\
  -  w^{+}_{1,0}[N_{1,0}]P(N_{1,0},N_{2,1}...,t) \nonumber \\
  +  w^{-}_{3,2}[N_{3,2}+1]P(N_{1,0}-1,N_{2,1}-1,N_{3,2}+1...,t) \nonumber \\
  +  w^{+}_{1,0}[N_{1,0}+2]P(N_{1,0}+2, N_{2,1} -1, N_{3,2}...,t) \nonumber \\
  - (w^{+}_{2,1}[N_{1,0},N_{2,1}] + w^{-}_{2,1}[N_{2,1}])P(N_{1,0},N_{2,1}...,t) \nonumber \\
  +  \sum_{n=3}^{N}\{w^{-}_{n+1,n}[N_{n+1,n}+1]P(N_{1,0} -1,...,N_{n,n-1}+1, N_{n+1,n}+1...,t)\} \nonumber \\
  +  w^{+}_{n-1,n-2}[N_{1,0}+1,N_{n-1,n-2}+1]P(N_{1,0}+1,...,N_{n-1,n-2}+1,N_{n,n-1}-1...,t) \nonumber \\
  -  (w^{+}_{n,n-1}[N_{1,0},N_{n,n-1}] + w^{-}_{n,n-1}[N_{n,n-1}])P(N_{1,0},...,t)
\end{multline}
The mean number of groups of size $(n,n-1)$ in case of incremental growth follows from: 
\begin{align}
  \mean{N_{n,n-1}(t)} = \sum_{\pmb{N}_{i}} N_{n,n-1}P(\pmb{N}_{i},t)
\end{align}
where $\pmb{N}_{i}$ refers to group distributions which have a total number of $N$ agents.
Using the master equation and the transition probabilities, we can write for the time-dependent change:
\begin{align}
  \label{eq:2}
  \frac{d\mean{N_{n,n-1}}}{dt} = \mean{-w^{-}_{n,n-1}(N_{n,n-1}) - w^{+}_{n,n-1}(n_{0}N_{n}) + w^{-}_{n+1,n}(N_{n+1,n}) + w^{+}_{n-1,n-2}(n_{0}N_{n-1,n})}
\end{align}
This equation holds only for $n\geq 2$.
For $n=1$ we have to take into account both the probability for any group to shrink and grow, as well as the probability for two agents to form a group or break apart. We refrain from writing out this longer equation here. Also, we will drop the arguments of the transition rates to lighten the notation.

By taking a Taylor expansion for the transition rates, the following Fokker-Plank equation (only dependent on $N_{n,n-1}$ and $n_{0}$ now) can be found \citep{Book-Ulbricht1988}: 
\begin{equation}
  \frac{d\mean{N_{n,n-1}}}{dt} = - \mean{\frac{\partial }{\partial n}\left\{ w^{+}_{n,n-1} - w^{-}_{n} \right\} } + \frac{1}{2}\mean{\frac{\partial^{2} }{\partial n^{2}}\left\{ w^{+}_{n,n-1} + w^{-}_{n} \right\} }
\end{equation}
We only consider the first part of this equation, neglecting  fluctuations in the system.
Using our transition rates and assuming $\mean{N_{n,n-1}n_{0}} = \mean{N_{n,n-1}}\mean{n_{0}}$, the first part can be written as follows:
\begin{align}
  \label{eq:6}
  \frac{d \mean{N_{n,n-1}}}{dt} = - \frac{\partial }{\partial n}  \left\{\frac{2m_{0}}{N^{2}}\mean{n}\mean{N_{n,n-1}}\mean{n_{0}} - \frac{\mean{n}\mean{N_{n,n-1}}}{N}\exp{\beta \frac{\hat{\rho}}{\rho}}\right\}
\end{align}
This can be transformed into a continuity equation, which  describes the in- and outflow of agents into the different groups:
\begin{align}\label{eq:4}
  \frac{d \mean{N_{n,n-1}}}{dt} + \nabla \cdot [\mean{N_{n,n-1}(t)}\cdot \dot{n}] = 0
\end{align}
This finally leads to the change in average group size $n$ which is discussed in the main part of the article already:
\begin{align}
  \label{eq:9}
  \frac{d \mean{n}}{dt} &= \frac{2m_{0}}{N^{2}}\mean{n}\mean{n_{0}} - \frac{\mean{n}}{N}\exp{\beta \frac{\hat{\rho}}{\rho}}\nonumber \\
                        &= \frac{\mean{y_{0}(t)}\mean{n}}{N} - \frac{\mean{n}}{N}\exp{\beta \frac{\hat{\rho}}{\rho}}\nonumber \\
  &=  \frac{\beta \hat\rho \mean{n}}{N}\left[ \frac{1}{\mean{\rho^{cr}_{n}(t)}} - \frac{1}{\mean{\rho(t)}}\right]
\end{align}

\end{appendix}

\small \setlength{\bibsep}{1pt}

\end{document}